\documentclass[twocolumn]{aastex631}
\usepackage[T1]{fontenc}
\usepackage{amsmath}
\usepackage{amssymb}

\usepackage{hyperref} 
\hypersetup{linkcolor=red,citecolor=blue,filecolor=cyan,urlcolor=blue}
\usepackage{graphicx}
\usepackage{amsmath}
\usepackage{bm}
\usepackage{ulem}

\newcommand{\rhobaryon}{\rho_{\text{B}}}

\newcommand{\gcc}{g cm$^{-3}$}
\newcommand{\gcs}{g cm$^{-2}$}
\newcommand{\mdotedd}{\dot{M}_\mathrm{Edd}}

\shorttitle{Thermonuclear Heating of Accreting Neutron Stars}
\shortauthors{Nava-Callejas, Page, \& Cavecchi}

\begin{document}

\title{Thermonuclear Heating of Accreting Neutron Stars}

\author[0000-0003-2334-6947]{Mart\'in Nava-Callejas}
\affiliation{Institute de Astronomie et Astrophysique, Université Libre de Bruxelles, 1050 Bruxelles, Belgique\\
Instituto de Astronom\'ia, Universidad Nacional Aut\'onoma de M\'exico, Ciudad de  M\'exico, CDMX 04510, Mexico}
\email{martin.javier.nava.callejas@ulb.be;\\ mnava@astro.unam.mx}

\author[0000-0003-2498-4326]{Dany Page}
\affiliation{Instituto de Astronom\'ia, Universidad Nacional Aut\'onoma de M\'exico, Ciudad de  M\'exico, CDMX 04510, Mexico}
\email{page@astro.unam.mx}

\author[0000-0002-6447-3603]{Yuri Cavecchi}
\affiliation{Departament de Fis\'{i}ca, EEBE, Universitat Polit\`ecnica de Catalunya, Av. Eduard Maristany 16, 08019 Barcelona, Spain}
\email{yuri.cavecchi@upc.edu}

\begin{abstract}

We describe a new method to incorporate thermonuclear heating in the envelope of accreting neutron star into long term simulations of their thermal evolution. 
We obtain boundary conditions for the heat exchange between the envelope and the crust based on stationary models which include nuclear burning and validate these values comparing to the results of the time-dependent code \texttt{MESA}.
These simple boundary conditions allow us to explore a large parameter space. 
We quantify the amount of heat flowing from the envelope into the crust, or viceversa, depending on the mass accretion rate, outburst duration and duty cycle, and especially crust/core physical parameters such as impurities, crustal heating, and neutrino cooling rate. 

\end{abstract}


\section{Introduction}

The heating of neutron stars undergoing accretion from a companion in a binary system has been a long and ``hot'' issue.
Gravitational energy, of the order of 100-200 
MeV per accreted baryon (denoted as ``MeV baryon$^{-1}$'' hereafter), 
released as heat at the surface when the accreting matter hits the neutron star leads to surface temperatures of the order of ten to twenty millions kelvins\footnote{
The surface temperature is limited by the Eddington luminosity, $L_\mathrm{Edd} \sim 10^{38}$ erg s$^{-1}$ \citep{Shapiro:1983wz}, implying a temperature of the order of $T_\mathrm{Edd} \sim 2 \times 10^7$ K. 
The corresponding limiting mass accretion rate is $\mdotedd \sim 2 \times 10^{-8} \, M_\odot \, \mathrm{yr}^{-1} \sim 10^{18}$ g s$^{-1}$, or 
$\dot{m}_\mathrm{Edd} = \mdotedd/4\pi R^2 \sim 10^5$ g s$^{-1}$ cm$^{-2}$.}.
This energy, however, is rapidly emitted by photons and does not penetrate into the star given that the underlying layers - just a few meters below - are much hotter than the surface as matter there undergoes thermonuclear reactions that rapidly raise the temperature well above $10^7$ K (see, e.g., \citealt{1998ASIC..515..419B} for simple analytical models of the outer layers of an accreting neutron star and their strong temperature gradients).

At densities $\sim 10^6$ \gcc, as a consequence of hydrogen and helium thermonuclear burning - either stable or unstable - the temperature rises up to about $10^{8}$ K (see e.g. \citealt{1981ApJS...45..389W, 1998ASIC..515..419B, 1999ApJ...524.1014S, 2000ApJ...544..453C, 2024arXiv240313994N}). 
In pioneer works studying the whole stellar temperature profile,  \citet{1984PASJ...36..199H} and \citet{1984ApJ...278..813F,Fujimoto1987op} identified two possible configurations:
in the case of strong core neutrino emission the stellar interior is colder than the burning envelope and heat flows from the envelope into the interior, what they call the ``watershed'' effect, while in the opposite case of a hotter interior the heat flow is from the interior into the envelope.
The first configuration has, however, received little attention while the second one has been studied more extensively due to the presence of further heating mechanisms identified in the crust.

As a consequence of mass accretion, deeper layers are compressed
 inducing further reactions and thus additional energy sources. 
 For instance \citet{1979PThPh..62..957S} and \citet{Bisnovatyi1979} proposed the presence of electron captures, neutron emissions, and pycnonuclear fusions in the crust, all collectively dubbed as ``deep crustal heating'' by \citet{Brown1998aa}.
Most studies of these mechanisms have shown that between one and two MeV baryon$^{-1}$ are generated in the crust, with pycnonuclear fusions releasing the majority of energy in the inner crust (see, e.g, \citealt{
1990A&A...227..431H,Haensel2003af,2008A&A...480..459H,Fortin2018,Gupta2007,Gupta2008,Lau2018aa}). 
In contrast, the competing scenario of \citet{Gusakov2020av} obtains a much smaller deep crustal heating estimate, $\sim 0.6$ MeV baryon$^{-1}$ \citep{Gusakov2021af,Shchechilin2021de,Shchechilin2022ty,Potekhin2023fg}.

Observations of the thermal relaxation of neutron stars in systems going into quiescence after a long phase of accretion have opened the possibility to literally map the internal temperature profile, and infer the location of the energy sources, through the time dependence of the cooling. 
Deep crustal heating by itself was found insufficient to explain the observed high temperatures and the inclusion of an extra energy source at lower densities, dubbed ``shallow heating'', was needed to improve these observation-theory comparisons \citep{Brown:2009aa}. 
Detailed studies of many cooling systems \citep{Brown:2009aa,Page2013ty,Degenaar2014io,Turlione2015rt,Degenaar2015nm,Merritt2016,Parikh2018kl,Parikh2019,Ootes2019rt,Degenaar2021bn} have found that this extra shallow source must usually inject an energy of the order of one to three MeV baryon$^{-1}$.
In one extreme case, the system MAXI J0556-332 \citep{Homan_2014}, a much larger amount of extra heating, between 5 to 15 MeV baryon$^{-1}$, was needed 
\citep{Deibel_2015,Parikh:2017uj}, but \cite{Page_2022} 
showed that most of this energy can be accounted for by a deep explosion, that they called  a ``hyperburst'', which occurred during the first observed accretion outburst and that a smaller amount of shallow heating, about half an MeV baryon$^{-1}$, was sufficient to interpret the observations of the next three outbursts.

Similarly, an extra source of heat in the outer layers of the crust has also been advocated in order to explain observations of Type I X-ray bursts and their related companions, the superbursts. 
Type I X-ray bursts are thermonuclear explosions that burn the accreted material, composed mainly of hydrogen, helium and traces of metals, at a column depth of $y \sim 10^8$ \gcs, corresponding to densities $\rho \sim 10^6$ \gcc (e.g. \citealt{1998ASIC..515..419B,Strohmayer:2003aa,Galloway:2021aa}). 
Superbursts \citep{cornelisse2000}, on the other hand, are thought to be triggered by carbon ignition at a much deeper layer, $y \sim 10^{12}$ \gcc\ (or $\rho \sim 10^9$ \gcc), the carbon having been produced by the previous burning in shallower layers \citep[e.g.][]{cumming2001,cumming2006,keek2011,keek2012,2016MNRAS.456L..11K}. 
Observationally, in most cases the unstable burning behind the Type I bursts stabilizes when the mass accretion rate raise above  $0.1-0.3 \, \mdotedd$
\citep{Cornelisse:2003aa,Watts:2007aa,Galloway:2008aa,Galloway:2021aa}, while theoretical models predict stabilization at $\dot{M} \sim \mdotedd$ \citep{1998ASIC..515..419B,2007ApJ...665.1311H,Fisker:2007aa}. 
In order to better match the theoretical predictions with the observations, an extra source of heat flux from the bottom layers has been invoked in several works  \citep[e.g.][]{keek2009,Zamfir:2014aa,2016MNRAS.456L..11K}. 
In the case of superburst, the presence of an extra heating source allows to explain certain observed properties such as ignition depth and recurrence time \citep[e.g.][]{cumming2006,keek2008,keek2011,meisel2024}. 
It is natural to consider that this extra heat source is related to the shallow heating described above. 

Several mechanisms that could contribute to this shallow heating have been examined. For instance, electron captures (beyond those in the basic deep crustal heating scenarios) and pycnonuclear fusions of neutron-rich isotopes of carbon and oxygen have been proposed \citep{Horowitz2008rt,Chamel2021we} since they could release $\sim 1$ MeV baryon$^{-1}$ at densities between $10^{10}$ to $10^{11}$ g cm$^{-3}$, depending on the chemical composition of the ashes from Type I X-ray bursts. Other, complementary, scenarios consider the effects of hydro- or magnetohydro-dynamics: matter falling from the inner accretion disc onto the neutron star is rotating much faster than the star itself so that not all the kinetic energy is released as heat but rather some differential rotation is maintained, leading to turbulent breaking at the interface of the ocean with the solid crust which could deposit large amounts of energy in these layers \citep{Inogamov1999,Inogamov2010}; alternatively, the effects of rotation, viscosity and instabilities  \citep{piro2007,keek2009} could also release extra heat. Less efficient processes could also contribute, such as compositionally driven convection that can provide about $\approx 0.2$ MeV baryon$^{-1}$\citep{Medin2011,Medin2014,Medin2015}.

During our study of stationary accreting neutron star envelopes \citep{2024arXiv240313994N}, we found that it may be possible to have accreting envelope models with a temperature inversion resulting in a negative luminosity 
$L_\mathrm{b}$ at its base and energy flowing into the crust from the envelope.
Our purpose in the present work is to study such models in details and consider their implications for the heating of accreting neutron stars. 
As mentioned above, such solutions had already been found many years ago in fully time-dependent simulations \citep{1984PASJ...36..199H,1984ApJ...278..813F, Fujimoto1987op} albeit with the simplification that they only included thermonuclear burning of H through the CNO cycle and no $^{4}$He burning. 
Later, \citet{Miralda-Escude1990wd} and \citet{Zdunik1992bn} presented similar but stationary models including the deep crustal heating, and also found solutions with temperature inversion within the envelope confirming the effect of enhanced core neutrino emission. More recently, \citet{Dohi2021, Dohi2022yu} using their neutron star evolutionary code \citep{Dohi:2020aa} obtained comparable results, with the major improvements that they performed time-dependent calculations of the whole neutron star, envelope and interior, they included hydrogen burning through the rp-process \citep{1981ApJS...45..389W}, as well as the explosive helium burning, but they focused on the analysis of X-ray bursts. 
Their models, however, could only cover a few days of evolution of the star.

Time-dependent models of the nuclear burning in the neutron star envelope are extremely CPU-time consuming and typically only allow the study of few models.
Our purpose is to perform a \textit{wide} exploration of the parameter space involved in the issue of the heating of accreting neutron stars from the thermonuclear burning in the envelope.
In order to be able to do this, in Section \ref{sec:stat_td_envelopes} we first consider stationary envelopes that allow us to generate numerous models covering the needed range of mass accretion rates and internal temperatures. Next, we present time-dependent envelope models, calculated with the public stellar evolution code \texttt{MESA}. Comparing our stationary models with these detailed time-dependent models we conclude that stationary models are an acceptable first approximation and that the occurrence of X-ray bursts does not significantly alter the long term structure of the envelope, a point that was already postulated by \citet{1984ApJ...278..813F}.
In Section \ref{sec:Boundary} we employ our stationary models to generate boundary conditions which we then implement in our neutron star evolution code \texttt{NSCool} in Section \ref{sec:Persistent_Accretion}.
We are, hence, able to fully explore the combined effects of the core neutrino emission and the mass accretion rate as well as the possible shallow heating, in addition to the deep crustal heating, and the uncertainty on the crust thermal conductivity. 
This allows us to obtain a detailed description of the conditions allowing leakage of thermonuclear energy from the envelope into the neutron star interior for the case of persistent accretion.
Finally, in Section \ref{sec:Transient_Accretion} we present a first look at the same issue but in the case of transiently accreting sources.
This last issue is more complicated and we can only study a generic scenario leaving the study of individual objects for future work.
 We present a discussion of our results and conclusions in Section \ref{sec:Conclusions}.
 
\section{Stationary and Time-Dependent Accreting Envelopes}
\label{sec:stat_td_envelopes}

\begin{figure*}
\begin{center}
\includegraphics[width=0.95\linewidth]{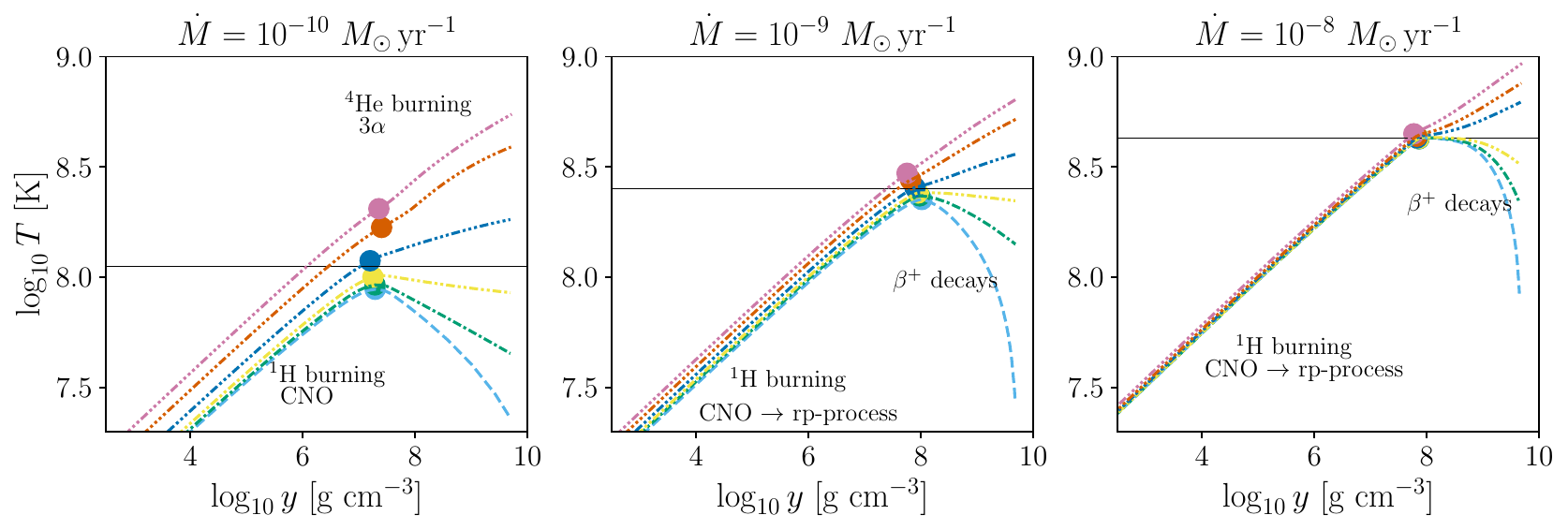}
\end{center}
 \caption{Examples of temperature profiles in stationary envelope models at three selected mass accretion rates, as marked in the upper part of the three panels. 
The horizontal lines denote the temperature $T_{\text{b}}$ at which $L_\text{b}=0$.
 Big dots mark the point where nuclear energy generation reaches its maximum:
 at lower densities $^1$H burning dominates, at higher densities $^4$He burning makes a significant contribution for 
 $\dot{M} =10^{-10} \, M_\odot$ yr$^{-1}$ while at the higher $\dot{M}$ energy generation by the $\beta^+$ decays becomes dominant due to the large amount of heavy nuclei produced by the rp-process.
Ample details of these phases of nuclear burning can be found in 
\citet{2024arXiv240313994N}.
 }
\label{fig:T-prof}
\end{figure*}

We first present here models of the envelope of an accreting neutron star in stationary state,
employing our stationary code described in \citet{2024arXiv240313994N}.
Our nuclear network, \texttt{net380}, contains 380 isotopes and includes H burning (through the pp-chain and the CNO cycle as well as the subsequent reactions of the rp-process up to nuclei of mass $A=107$), He burning (through 3$\alpha$ and $\alpha$ captures), and
$^{12}$C--$^{12}$C, $^{12}$C--$^{16}$O, and $^{16}$O--$^{16}$O fusions.
For the radiative opacity, we opted for implementing the analytic fits of \citet{1999ApJ...524.1014S} and \cite{2017ApJ...835..119P} for free-free and electron-scattering processes respectively. 
Their sum was corrected employing the factor from \cite{2001A&A...374..213P}. 
For the conductive opacities we consider both \citet{1999ApJ...524.1014S} and \cite{2006PhRvD..74d3004S}.

The calculations start at the photosphere, at an optical depth of $2/3$, with a chosen value of its effective temperature $T_\mathrm{eff}$, from which the surface density $\rho_\mathrm{s}$ and pressure $P_\mathrm{s}$ are deduced using an Eddington-type condition. 
More involved boundary conditions are possible, but they have little effect on the deeper layers as the initial condition is rapidly forgotten \citep{Miralda-Escude1990wd,2012sse..book.....K}.
From this point on, the envelope equations are integrated inward until the density reaches a chosen boundary value $\rho_\mathrm{b}$.
We chose $\rho_\mathrm{b} = 10^7$ g cm$^{-3}$ since the generation of nuclear energy past this value is negligible.
Such boundary condition, $\rho = 10^{7}$ g cm$^{-3}$, is a typical value in the literature of accreting neutron stars, both in stationary (e.g. \cite{1999ApJ...524.1014S}) or time-dependent simulations \cite{2007ApJ...665.1311H}.
The structure of the envelope in stationary condition depends on the mass accretion rate per unit area $\dot{m}=\dot{M}/4\pi R^2$ through the change in chemical composition when matter is being pushed from a depth $y$ to $y+dy$ and this change is proportional to the time $dt=dy/\dot{m}$ it takes for matter to move a depth $dy$.
Moreover, when passing from the layer $y$ to $y+dy$ the luminosity $L(y)$ is reduced by the amount of nuclear and gravitational energy generated within this layer, starting from $L = 4\pi R^2 \sigma_{SB} T_\mathrm{eff}^4$ at the photosphere. 
When choosing a low enough $T_\mathrm{eff}$ it is possible for $L(y)$ to turn negative.

In Fig.~\ref{fig:T-prof} we illustrate the temperature profiles of three families of stationary envelopes, accreting a solar-like chemical composition, at three different rates.
We consider a star of mass $M = 1.4 M_{\odot}$ and radius $R = 11.56\, \mathrm{km}$ - corresponding to an APR-core EOS star \citep{1998PhRvC..58.1804A}.
In all three cases we observe some envelopes where $T(y)$ reaches a maximum and then decreases, resulting in negative luminosities and a leakage of thermonuclear energy from the burning envelope into the neutron star interior.
The horizontal lines denote the temperature $T_{\text{b}}$ at which $L_\text{b}=0$, a critical value between envelopes with $L_\text{b}$ positive versus negative.
Naturally this critical $T_{\text{b}}$ increases with increasing $\dot{M}$.

In order to explore the energy leakage in a fully time-dependent context, as well as to make a comparison with the stationary state predictions, we simulated the evolution of neutron star envelopes employing the publicly available code \texttt{MESA} v.15140 \citep{Paxton:2011aa,Paxton:2013aa,Paxton:2015aa,Paxton:2018aa,Paxton:2019aa,Jermyn:2023aa}. 
As initial condition we implemented a fiducial neutron star envelope of pure $^{56}$Fe. 
This envelope sits on top of an inner core, for which we adopt a mass $M = 1.4\ M_{\odot}$ and a radius $R = 11.56$ km, that only provides the inner boundary condition for the \texttt{MESA} evolution. 
For the radiative opacity, also for \texttt{MESA} we used the analytic fits of \cite{1999ApJ...524.1014S} and \cite{2017ApJ...835..119P} for free-free and electron-scattering processes, respectively, and the correction factor of  \cite{2001A&A...374..213P}.
For the conductive opacities we employed the tables provided by \texttt{MESA}.
To avoid exceedingly long computing times, instead of the \texttt{net380} nuclear network we adopted a shorter one, \texttt{Approx149}, which is a slightly improved version of the \texttt{Approx140} network of \citet{2024arXiv241109843N} adding 9 nuclides: $^{56}$Fe, $^{61}$Cu, $^{22-23}$Na, $^{23}$Mg, $^{26-27}$Al, $^{27}$Si and $^{31}$P. 

Relativistic corrections in \texttt{MESA} are implemented by multiplying $G$ with $(1 - 2Gm/c^{2}R)^{-1}\left(1+ \frac{4\pi P r^{3}}{mc^{2}}\right)(1 + P/\rho c^{2})$. 
However, such scheme misses several $e^{\Lambda}$ factors in the temperature and luminosity equations (see Appendix~\ref{appendx:gr_corr_1}) 
and we, thus, omitted including these corrections.
For the present comparison among codes we adapted our stationary envelope code to be non-relativistic as well. 
To ensure nuclear burning and mass compression were the only heating sources - as well as to accelerate the numerical simulation - we disabled convective and thermohaline effects in \texttt{MESA}.

In envelope modeling, \texttt{MESA} takes the luminosity entering the base, $L_\mathrm{b}$, as its inner boundary condition and keeps it constant during the whole simulation.
In a more realistic situation, $L_\mathrm{b}$ should evolve due to the evolution of the whole neutron star.
In a first attempt, and in order to minimize the effect of the inner boundary condition, we put it at relatively high density, $\rho_b = 2 \times 10^9$ \gcc, starting with $T_{b} = 10^8$ K and fixing 
$L_b$ at 0.25 $L_\odot$.
We consider four different accretion rates, $10^{-9}$, $3\times 10^{-9}$, $10^{-8}$ and $3\times 10^{-8 }$ in units of $M_{\odot}\, \mathrm{yr}^{-1}$.

\begin{figure*}
\begin{center}
\includegraphics[width=0.95\linewidth]{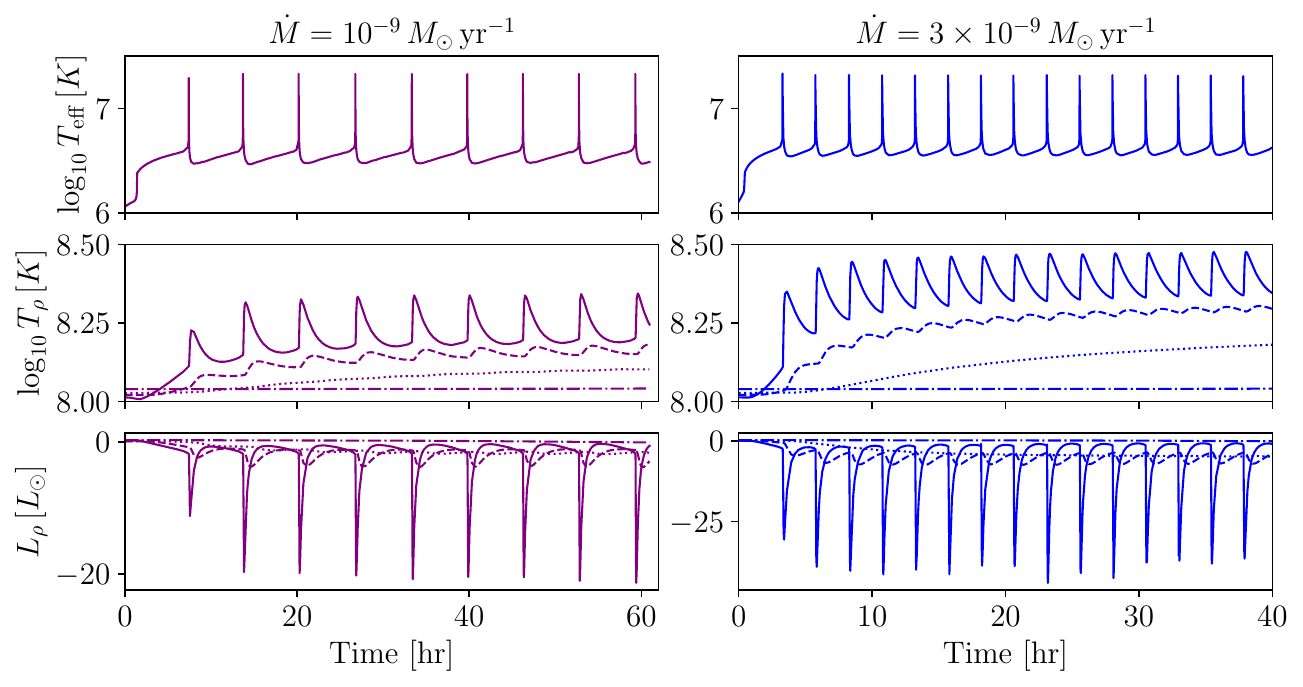}
\hspace{\stretch{1}}
\includegraphics[width=0.95\linewidth]{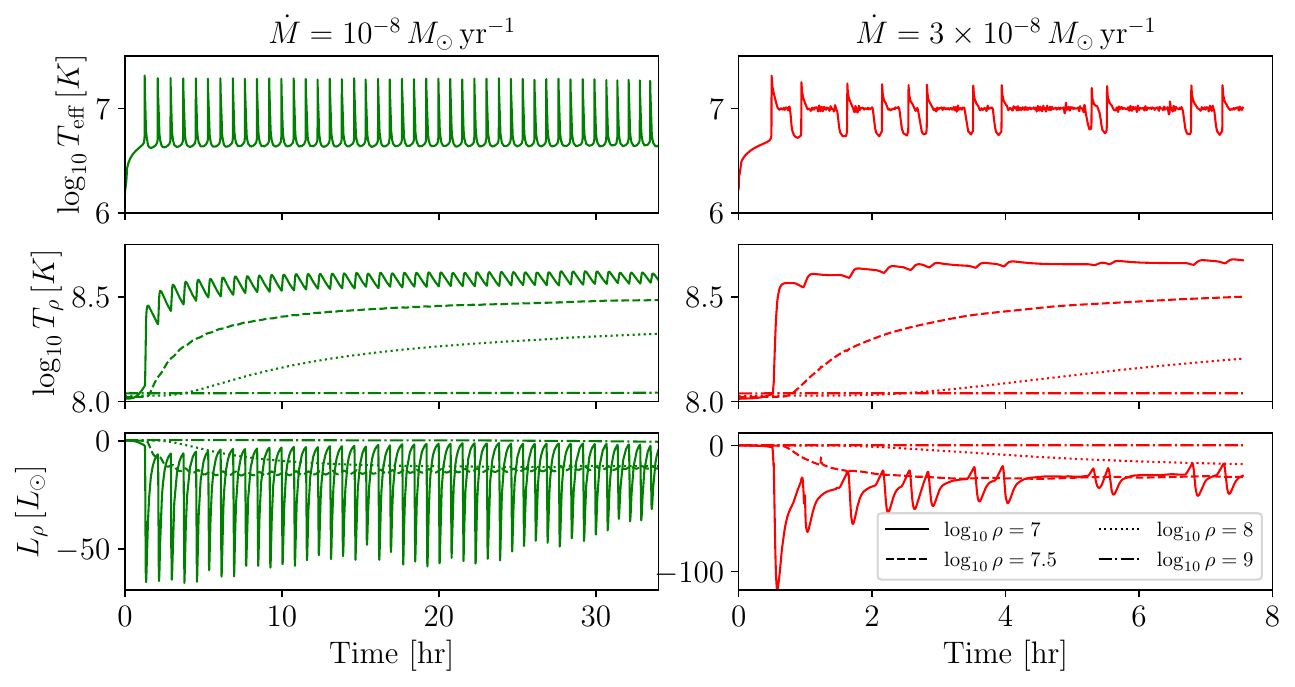}
\end{center}
 \caption{Time evolution of \texttt{MESA} models of accreting envelopes with four different mass accretion rates, as indicated in each panel.
 In each panel the upper frames present the evolution of $T_\mathrm{eff}$, the central frames the temperatures $T_\rho$ at four different densities, as labeled in the lower right panel, and the lower frames the luminosity $L_\rho$ at the same densities.
 See text for details.
 }
\label{fig:mesa_different_rates}
\end{figure*}

In Fig.~\ref{fig:mesa_different_rates} we illustrate three properties of these envelope models as a function of time: the effective temperature, and the luminosity and internal temperature at four different densities.
In the four cases we observe a similar trend: given the assumed low value of $T_\mathrm{b}$, energy progressively flows from the envelope towards the crust. 
In particular, we observe a large deposition of energy at $\log_{10}\rho$ [\gcc] = 7 and 8, between 0.1 and 0.25 MeV baryon$^{-1}$. 
At $\log_{10}\rho$ [\gcc] = 9, on the other hand, little effect is observed in both temperature and luminosity given the large amount of time, larger than the duration of our simulations,  required for heat to reach this density.
A common feature of the models with $\dot{M} < 3\times 10^{-8}$ $M_{\odot}$ yr$^{-1}$ is the oscillatory behavior in temperature and luminosity at $\log_{10}\rho$ [\gcc] = 7, following the trend of the effective temperature and as a consequence of the thermonuclear explosions reported by \texttt{MESA}. 
Such rapid variation can be explained by its proximity with the location of the ignition, at densities $\sim 10^{6}$ \gcc. 
At $\dot{M} = 3\times 10^{-8}$ $M_{\odot}$ yr$^{-1}$, on the other hand, temperature and luminosity oscillations at $\log_{10}\rho$ [\gcc]= 7 become less accentuated than in the other models, suggesting the nuclear burning is approaching the stable burning regime. 
Regardless of the rapid variations at $\log_{10}\rho$ [\gcc]= 7 for the temperature of the $\dot{M}< 3\times 10^{-8}$ $M_{\odot}$ yr$^{-1}$ envelopes, we see that their minima, just before the initiation of the next burst, follow a monotonically increasing trend, similar to what happens with their corresponding curves at $\log_{10}\rho$ [\gcc]= 7.5 and 8. 
For luminosity, however, this effect occurs between $\text{log}_{10} \rho$ [\gcc] = 7.5 and 8. We notice this trend is actually prominent at $\dot{M} = 3\times 10^{-8}$ $M_{\odot}$ yr$^{-1}$. 
Despite the presence of bursts, around $\log_{10}\rho = 7$ and 7.5 we observe the temperature to be converging towards an asymptotical value. This suggests that we could consider an average temperature and luminosity as a replacement for the fully time-dependent behavior to compare with the stationary models. 

\begin{figure}
\begin{center}
\includegraphics[width=0.99\linewidth]{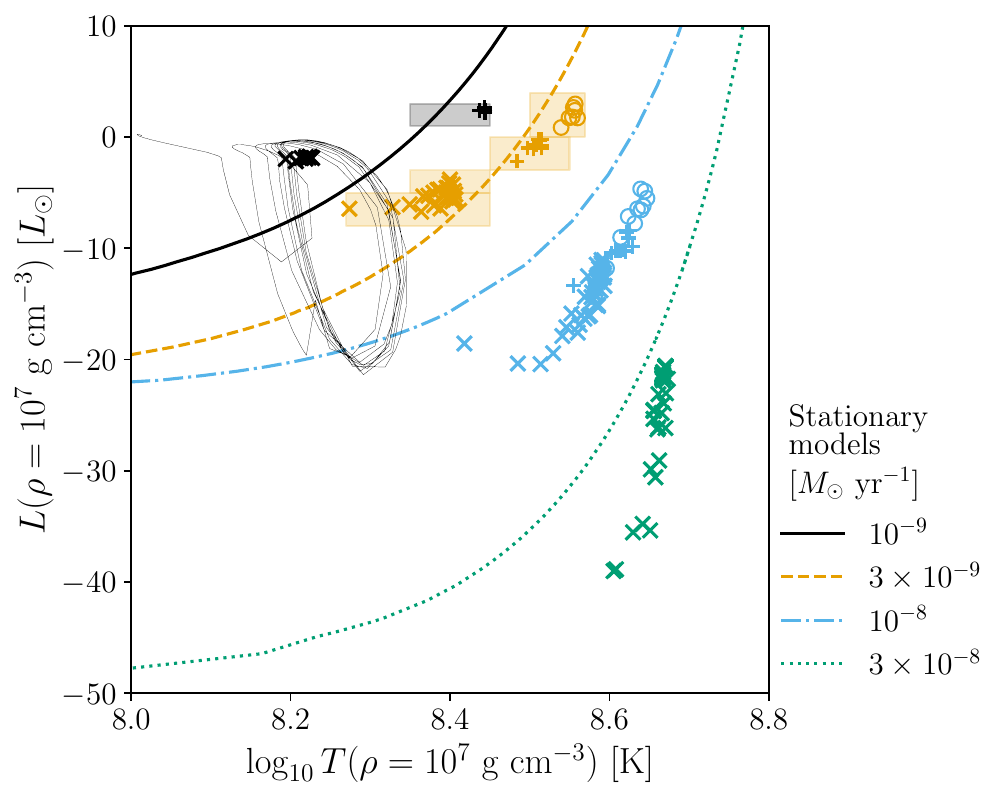}
\end{center}
\caption{
Comparison of the luminosity versus temperature, at $\rho = 10^7$ \gcc\ and four different accretion rates, of our stationary models, thick lines, and burst averages of our \texttt{MESA} time dependent models.
Crosses, $\times$, show averages over each of the bursts presented in Figure \ref{fig:mesa_different_rates}, 
with $L_b = 0.25 L_\odot$ and $T_b = 10^8$ K;
pluses, $+$, do the same for models with 
$L_b = 4 L_\odot$ and $T_b = 10^{8.6}$ K; 
and circles, $\bigcirc$,
for $L_b = 7 L_\odot$ and $T_b = 10^{8.7}$ K. The thin line shows the complete time evolution of $L$ versus $T$ for the model with $\dot{M} = 10^{-9} \, M_\odot$ yr$^{-1}$, from which the average values, crosses, are obtained.
Shaded regions are included to avoid ambiguities in the time dependent \& independent comparison.
In all cases $T$ at $10^7$ \gcc\ is increasing with time.
See text for details.
 }
\label{fig:mesa_vs_ss_average}
\end{figure}

We display the temperature and luminosity averages in Figure \ref{fig:mesa_vs_ss_average} for the same four mass accretion rates, together with complementary models at different base luminosities and temperatures.
For these averages we require a time interval: we employed the time interval between two successive peaks of effective temperature, i.e. a bursting cycle. The number of averages thus corresponds one-to-one with the number of simulated bursts and together they show a time evolutions similar to the stationary models curves.
Unfortunately, reaching the stationary states would require much more computing time and would, anyway, only explore a very small part of the $L_b$--$T_b$ parameter space.
Nevertheless these results show that our stationary models appear to be a reasonable approximation.

\begin{figure}
\begin{center}
\includegraphics[width=0.99\linewidth]{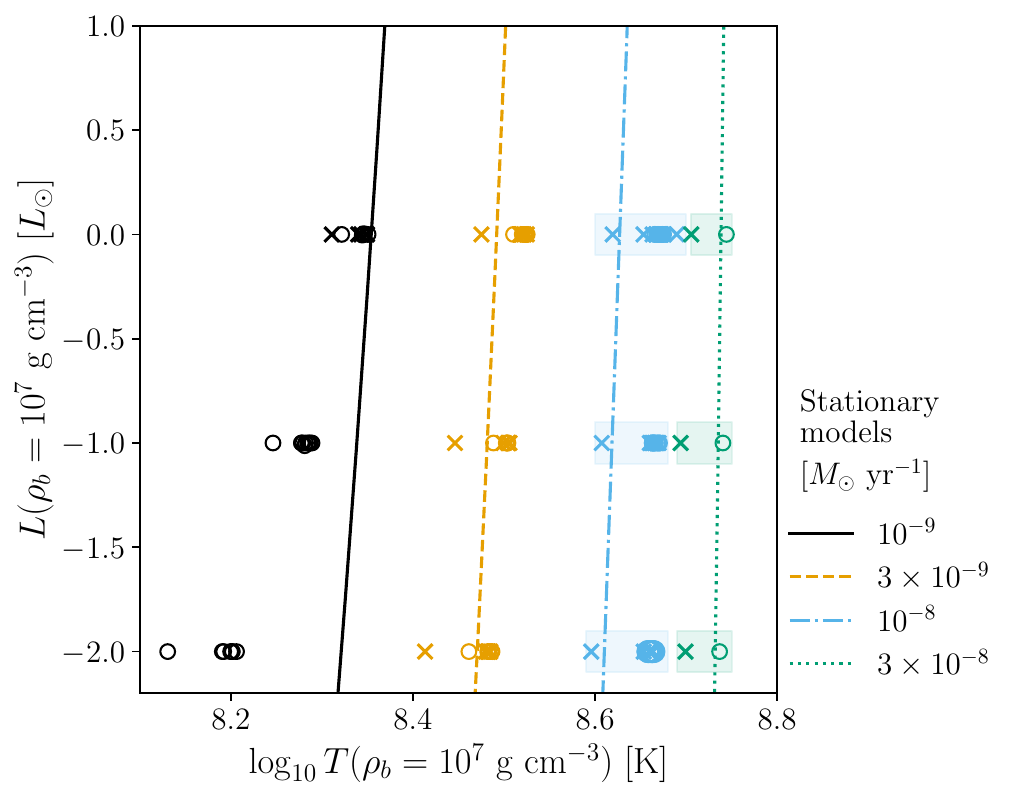}
\end{center}
 \caption{Comparison of the luminosity versus temperature, at four different accretion rates, of our stationary models, lines, and burst averages of our \texttt{MESA} time dependent models with boundary at $\rho_b = 10^7$ \gcc.
Shaded regions are included to avoid ambiguities in the time dependent \& independent comparison.
Symbols have the same meaning as in Figure \ref{fig:mesa_vs_ss_average}.
 }
\label{fig:mesa_vs_ss_Lb=0}
\end{figure}

There is a second way of comparing $L$ and $T$ at $\rho= 10^7$ \gcc\ between the stationary models and the average values of time dependent \texttt{MESA} models by setting the boundary of the \texttt{MESA} models at this density, i.e., choosing $\rho_b = 10^7$ \gcc.
For each chosen value of $L_b$, $T_b$ is then solved for self-consistently by \texttt{MESA}.
We generated a series of models with various values of $L_b$ starting with either a ``hot'' ($T_{\text{eff}} = 5\times 10^{6}$ K) or ``cold'' ($T_{\text{eff}} = 10^{6}$ K) envelope, relaxed towards the desired base luminosity via the \texttt{MESA} command \texttt{relax\_L\_center}, and allowed them to evolve.
In Figure~\ref{fig:mesa_vs_ss_Lb=0} we show the results of these simulations.
It is noticeable that after a few bursts the average temperature between bursts has settled close to the stationary envelope values, for all the selected $L_b$'s at the selected accretion rates, regardless of whether they started cold or hot.
It is important to notice that the agreement is particularly good at the critical value of $L_b = 0$, a fact of importance for our studies presented below.

\section{Stationary Accreting Envelopes as Boundary Conditions for Modeling Accreting Neutron Star Interiors} 
\label{sec:Boundary}

\begin{figure*}
\begin{center}
\includegraphics[width=0.49\linewidth]{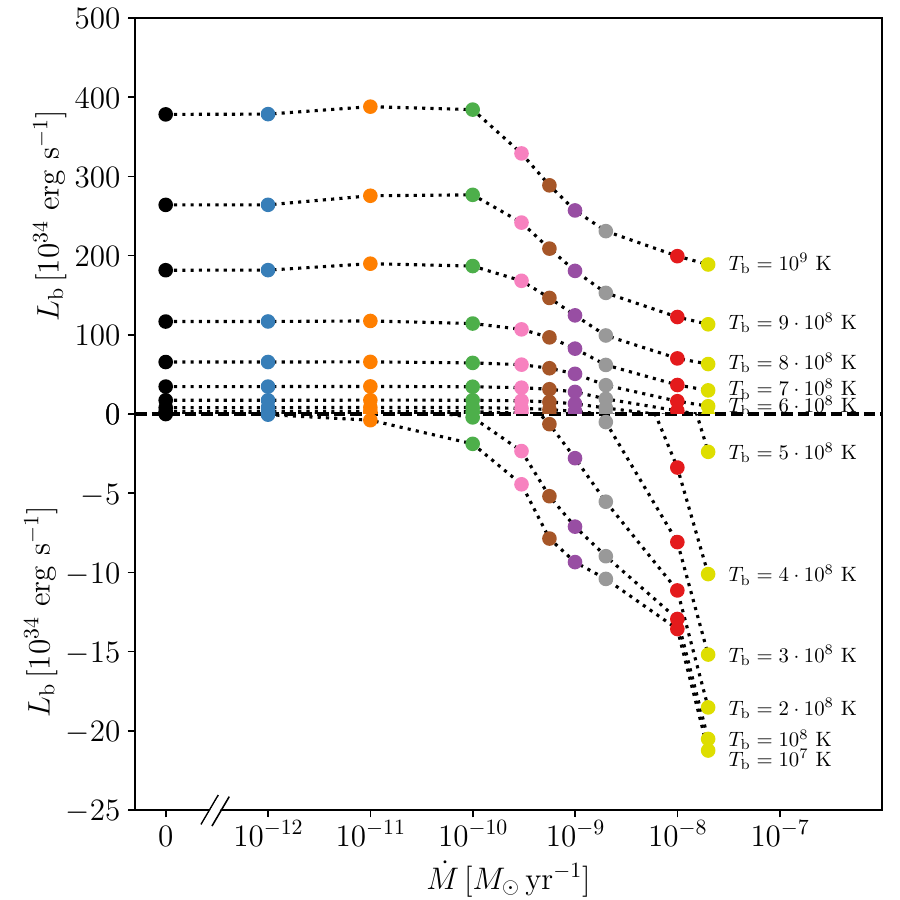}
\includegraphics[width=0.49\linewidth]{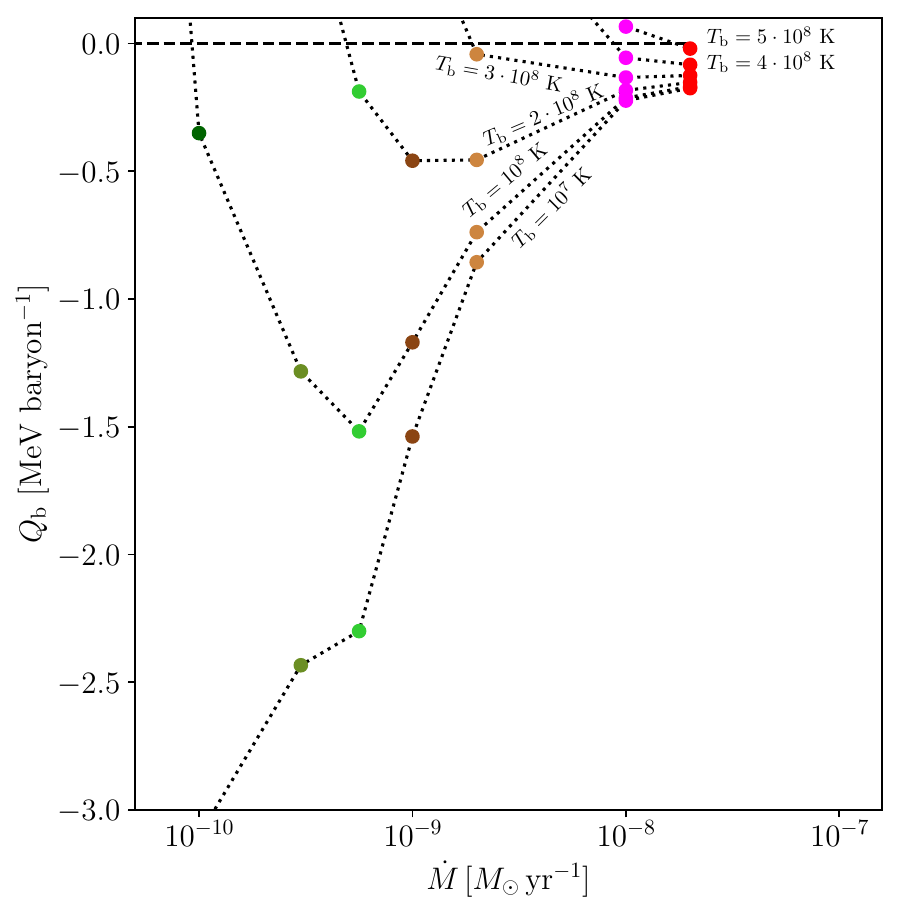}
\end{center}
 \caption{Left panel: envelope base luminosity, $L_\mathrm{b}$, versus mass accretion rate, $\dot{M}$, for several base temperatures, $T_\mathrm{b}$.
 Notice the change of $L_\mathrm{b}$ scale between the upper (positive $L_\mathrm{b}$) versus lower (negative $L_\mathrm{b}$) parts. Right panel: equivalent envelope base energy per baryon, $Q_\mathrm{b}$, versus mass accretion rate, $\dot{M}$, for several base temperatures, $T_\mathrm{b}$, corresponding to the $L_\mathrm{b}$ of the left panel with $L_\mathrm{b} <0$. A table of values of $L_\mathrm{b}$ interpolated on uniform grids of 
 $T_\mathrm{b}$ and $\log_{10} \dot{M}$
 from the ones used for this figure is
 available as a Table online.}
\label{fig:Lb-Tb}
\end{figure*}

When modeling the thermal evolution of a neutron star one generally does not solve the thermal evolution equations out to the photosphere, but one rather introduces an outer boundary, located at a fixed density $\rho_\mathrm{b}$, or a fixed pressure $P_\mathrm{b}$, where an appropriate boundary condition is imposed.
The latter is simply a relationship between the temperature, $T_\mathrm{b}$, and the luminosity, $L_\mathrm{b}$, at this outer boundary point (see, e.g., Appendix B in \citealt{Page_2004}), i.e., a Cauchy boundary condition.
Such relationship is obtained from envelope models as the ones presented for example by \citet{1983Gudmundsson} for envelopes made of pure iron or \citet{Potekhin:1997mn}, who generalized them to include accreted matter.
Our stationary models presented in the previous section provide us with a further generalization to the case of accreting envelopes that automatically include in their structure the energy generated by the thermonuclear burning.
For this purpose we employ our extensive \texttt{net380} network (see \citealt{2024arXiv240313994N}) and include general relativistic effects that were artificially excluded in the previous sections for comparison with \texttt{MESA}.

We show in the left panel of Figure \ref{fig:Lb-Tb} the results of a series of calculations for various mass accretion rates that can be used as such $T_\mathrm{b}-L_\mathrm{b}$ boundary conditions that become $\dot{M}$-dependent.
Notice that what is considered a base temperature $T_\mathrm{b}$ or base luminosity $L_\mathrm{b}$ when calculating an envelope models becomes an outer boundary temperature $T_\mathrm{b}$ or boundary luminosity $L_\mathrm{b}$ when applied for modeling the evolution of the stellar interior.
When $\dot{M} \ll 10^{-10} \, M_\odot$ yr$^{-1}$ the obtained luminosities are almost identical to the ones at $\dot{M} = 0$.

In the right panel of Figure \ref{fig:Lb-Tb} we focus on a part of this diagrams where negative luminosities are found, i.e., where heat flows {\it from the envelope into the crust}, and where we have converted $L_\mathrm{b}$ into an energy per accreted baryon $Q_\mathrm{b}$ through
\begin{equation}
L_\mathrm{b} =\frac{\dot{M}}{m_u} Q_\mathrm{b}
\label{Eq:Lb-Tb}
\end{equation}
where $m_u$ is the atomic mass unit.
We see that in the case of low stellar interior temperature, $\leq 10^8$ K, the amount of thermonuclear energy flowing into the star can be very significant, actually comparable to the amount of ``shallow heating'' that has been found necessary to invoke to interpret observations of crust relaxation in transiently accreting neutron stars which is of the order of 1 to 3 MeV per accreted baryon
\citep{Page2013ty,Degenaar2014io,Turlione2015rt,Degenaar2015nm,Merritt2016,Parikh2018kl,Parikh2019,Ootes2019rt,Degenaar2021bn,Page_2022}.

Notice that in our present models we have assumed a solar composition for the accreted matter and a neutron star of mass $1.4 \, M_\odot$ 
and radius of 11.56 km.
The effect of changing these values will be considered in a future paper.

\section{Modeling Persistently Accreting Sources with their Thermonuclear Burning.}
\label{sec:Persistent_Accretion}

We will here investigate in which conditions the above described envelope models actually result in thermonuclear energy flowing from the burning layers into the neutron star crust in realistic situations.
For this we employ an updated version of our neutron star cooling code \texttt{NSCool} \citep{NSCool} to simulate the evolution of a star until it has reached a stationary state under a constant mass accretion rate $\dot{M}$.
At that point the red-shifted core temperature, $\widetilde{T}$, is expected to be uniform due to the extremely high thermal conductivity of the core matter.

As heating source we apply the deep crustal heating as described by 
\cite{2008A&A...480..459H} that injects about 1.9 MeV baryon$^{-1}$, distributed in the crust\footnote{In the alternate model of \citet{Gusakov2020av} the energy released by the deep crustal heating is much smaller, of the order of 0.5 MeV \citep{Potekhin2023fg}. We considered this possibility but noticed it has only a very small effect on the impact of the crust heating from the envelope.}.
We also consider the effect of different amounts of shallow heating, $Q_\mathrm{sh}$, up to 3 MeV baryon$^{-1}$ that are in the range of most values found necessary to interpret observations of crust relaxation in transient sources (see, e.g., \citealt{Page2013ty,Degenaar2014io,Turlione2015rt,Degenaar2015nm,Merritt2016,Parikh2018kl,Parikh2019,Ootes2019rt,Degenaar2021bn,Page_2022})
and different densities, $\rho_\mathrm{sh}$, at which this energy is released, which we spread in the range $[\rho_\mathrm{sh}, 10 \rho_\mathrm{sh}]$.

For the cooling of the star we consider the two distinct scenarios of ``slow'' and ``fast'' neutrino emission
(see, e.g., \citealt{Yakovlev:2004iq,PAGE2006497,Potekhin2015ty}). The former are processes that involve five degenerate fermions and have an overall temperature dependence of $T^8$ while the second are processes involving 3 degenerate fermions with a $T^6$ behavior.
Instead of employing definite models of equation of state and their associated neutrino emission processes we prefer writing a generic neutrino luminosity $L_\nu$ as
\begin{equation}
L_\nu = N_8^\mathrm{slow} \cdot \widetilde{T}_8^8
\;\;\;\; \mathrm{or} \;\;\;\;
L_\nu = N_8^\mathrm{fast} \cdot \widetilde{T}_8^6
\label{Eq:L_nu}
\end{equation}
where $\widetilde{T}_8 = \widetilde{T}/ 10^8$ K.
These luminosities of isothermal stars have been studied in detail by \citet{Ofengeim2017wr}, see also \citet{2013MNRAS.432.2366W} and \citet{Page:2025aa}, who found values of $N_8^\mathrm{slow}$ ranging from $10^{31}$ to $10^{35}$ erg s$^{-1}$ and $N_8^\mathrm{fast}$ up to $10^{43}$ erg s$^{-1}$ when expressed as a function of the red-shifted core temperature $\widetilde{T}$.
In the presence of pairing in the core the fast neutrino emission can be dramatically reduced and so we consider minimal values of $N_8^\mathrm{fast}$ down to $10^{35}$ erg s$^{-1}$ to cover the whole range of possible values of $L_\nu$.
The other cooling agent, photon emission at the surface, is automatically taken into account by the outer boundary condition as described in the previous section.

Beside the heating and cooling agents we need to set the thermal conductivity $K$. 
The core conductivity is extremely high and of no concern as it will easily result in an isothermal core.
Of outmost importance, however, is the crust conductivity as it controls how the accretion heat will diffuse through it and eventually into the core or toward the surface.
In the crust the conductivity is strongly dominated by the electron contribution and we write it in terms of their collision frequency $\nu_\mathrm{e}$ as
\begin{equation}
K = \frac{c_\mathrm{V}(e) v_\mathrm{F}(e)^2}{3\nu_\mathrm{e}} 
\label{Eq:K}
\end{equation}
where $c_\mathrm{V}(e)$ is their specific heat, at constant volume, per unit volume and $v_\mathrm{F}(e)$ their Fermi velocity.
In the region where the nuclei form a liquid $\nu_\mathrm{e} = \nu_\mathrm{e-ion}$ which we calculate following \citet{Yakovlev1980sa}.
When nuclei solidify and crystallize we write 
$\nu_\mathrm{e} = \nu_\mathrm{e-ph} + \nu_\mathrm{e-imp}$ and follow \citet{Potekhin1999pl} for the electron-phonon scattering frequency $\nu_\mathrm{e-ph}$ and \citet{Yakovlev1980sa} for the electron-impurity scattering frequency $\nu_\mathrm{e-imp}$, the latter being the major uncertainty in the thermal conductivity of the crust.
This effect of impurities is parametrized by the ``impurity parameter'' $Q_\mathrm{imp}$ \citep{Flowers1976op,Yakovlev1980sa,Itoh1993er}, but
we emphasize that this treatment of impurity is highly uncertain and 
$Q_\mathrm{imp}$ should be considered more as a ``fudge factor''\footnote{
We notice that, approximately, $\nu_\mathrm{e-ph}$ is $\propto T$ while $\nu_\mathrm{e-imp}$ is temperature independent. Hence, the writing $\nu_\mathrm{e} = \nu_\mathrm{e-ph}+ \nu_\mathrm{e-imp}$ is a form of $T$-expansion in a physically motivated manner.} 
that mimics more complex processes (see, e.g., \citealt{Roggero2016wd,CaplanHorowitz2017}).
Its value can be empirically constrained in studies of crust relaxation in transient sources, as mentioned above, where values smaller than 10 were found necessary, while in the case of persistent sources we have no other choice than cover a wide range of values for an exploratory purpose.
We will, hence, consider values of $Q_\mathrm{imp}$ up to 100, but remember that much smaller values, less than 10, have been found empirically.

Considering the equation of state (EOS), at densities below $10^{10}$ g cm$^{-3}$ we apply the same procedure as in the stationary envelope models described above. 
At higher densities in the crust we follow \cite{2008A&A...480..459H} and \citet{Negele:1973tp}.
To have a complete neutron star model that can be evolved with \texttt{NSCool} we need a complete description of the core of the star for which
we apply the EOS of \citet{1998PhRvC..58.1804A} and base our modeling on a $1.4 \, M_\odot$ model with a 11.56 km radius.
Considering stars of different mass and radius will change the thickness of the crust and will have some impact on the results
and we leave discussing this issue for future works as we already have a large number of variables in our modeling.
However, by adjusting by hand the neutrino luminosity of the core, Equation (\ref{Eq:L_nu}), we take into account its major effect and mimick the evolution of stars of different masses.
 
Finally, we use the $\dot{M}$-dependent boundary condition presented in the previous section.
With this procedure we automatically include nuclear energy generation 
in the modeling, as it is part of the envelope models, and will dynamically find how it interacts with the thermal evolution of the stellar interior.
We are, thus, including three types of energy sources: the thermonuclear heating in the envelope, the shallow heating in the outer crust, and the deep crustal heating acting in the whole crust but mostly concentrated in the inner crust where the pycnonuclear reactions take place.

Having fixed all parameters of a given model, we run \texttt{NSCool} with constant mass accretion rate for a million years of stellar time, guaranteeing that the star has reached a steady state \citep{2001ApJ...548L.175C}.
This takes only a few seconds of CPU time per model and allows us to explore a large parameter space.

\subsection{Results.}
\label{sec:Persistent_Accretion_Results}

\begin{figure*}
\begin{center}
\includegraphics[trim={0 0 0 1.0cm},width=0.90\textwidth, angle=0]{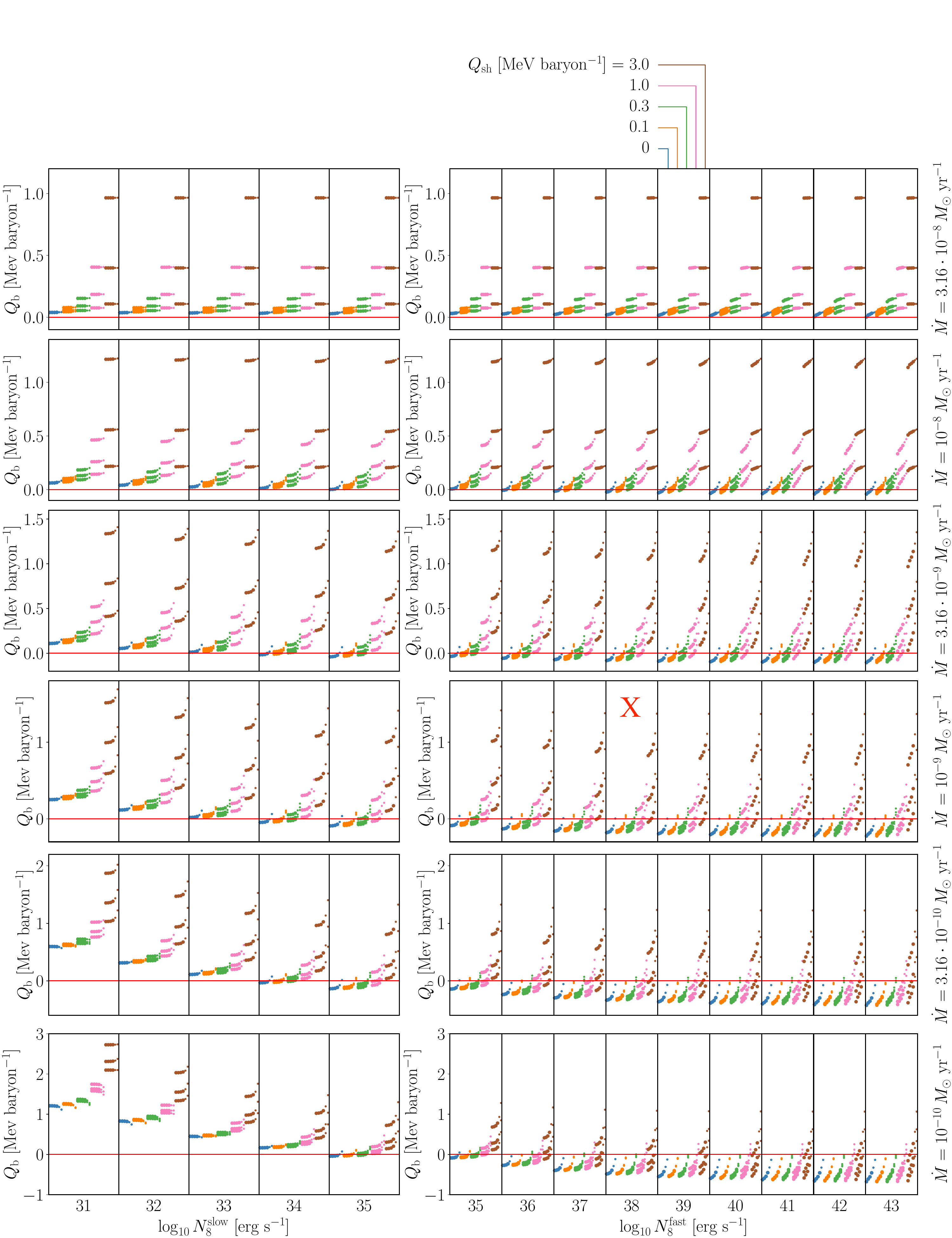}
\end{center}
 \caption{Energy transfer in persistently accreting models:
 energy per accreted baryon, $Q_\mathrm{b}$, transferred between the crust and the envelope as a function of the neutrino luminosity, $N^\mathrm{slow}_8$ for slow processes in the left panel and $N^\mathrm{fast}_8$ for fast processes in the right panel, which are, moreover, separated in six rows of frames according to the assumed mass accretion rate $\dot{M}$.
 In each frame with given values of $N^*_8$ and $\dot{M}$, the five vertical series of points correspond to five different amounts of shallow heating, as indicated at the top of the figure, and each series consists of 3 strings of 6 points according to three different values of $\rho_\mathrm{sh} = 10^8$, $10^9$, and $10^{10}$ g cm$^{-3}$ and six values of $Q_\mathrm{imp} =$ 0, 1, 3, 10, 30, and 100. 
 Details of a sample frame are described in Figure \ref{fig:analysis}.
 Heat flows from the envelope into the crust in models where $Q_\mathrm{b}<0$.
 }
\label{fig:Nu-Mdot}
\end{figure*}

\begin{figure*}
\begin{center}
\includegraphics[trim={0 0.2cm 0 0},width=0.40\textwidth]{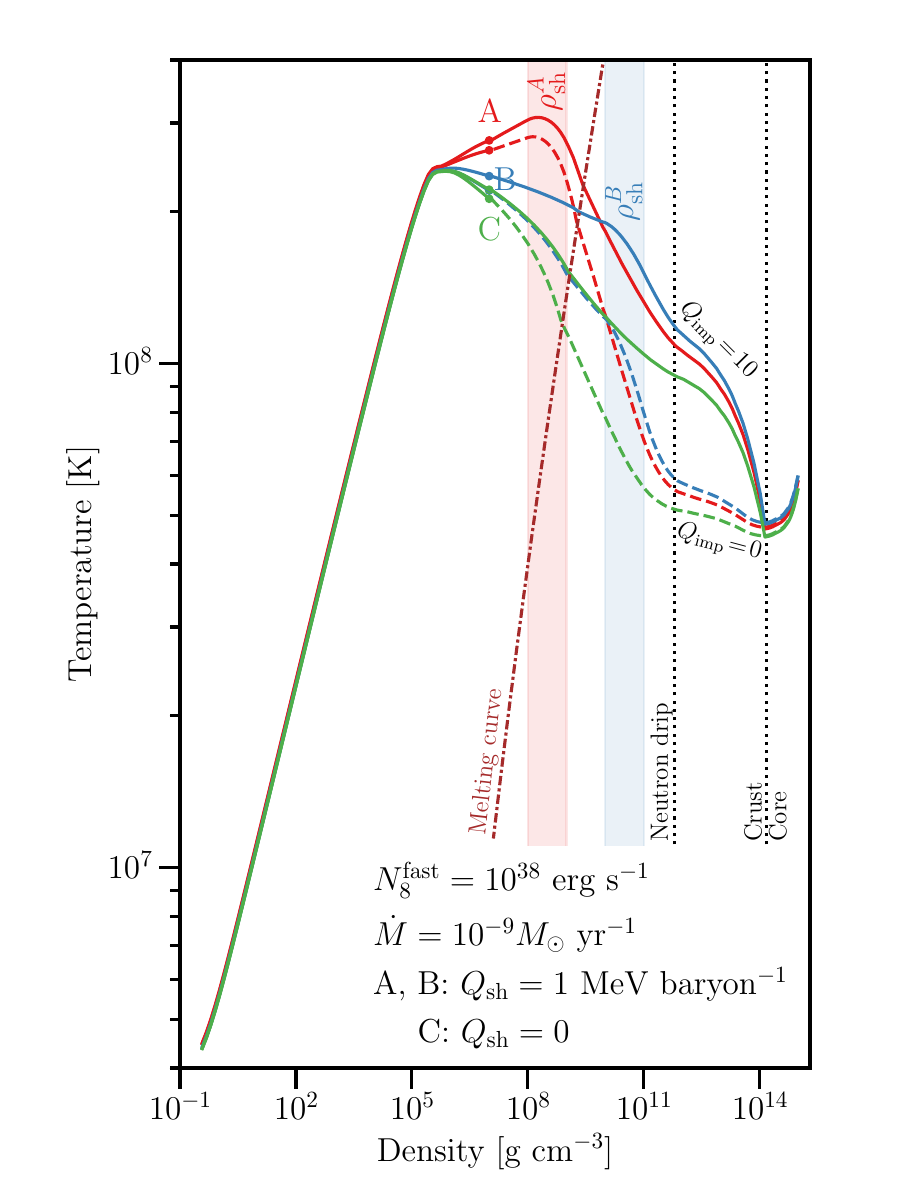}
\includegraphics[trim={0 2.0cm 0 0},width=0.30\textwidth]{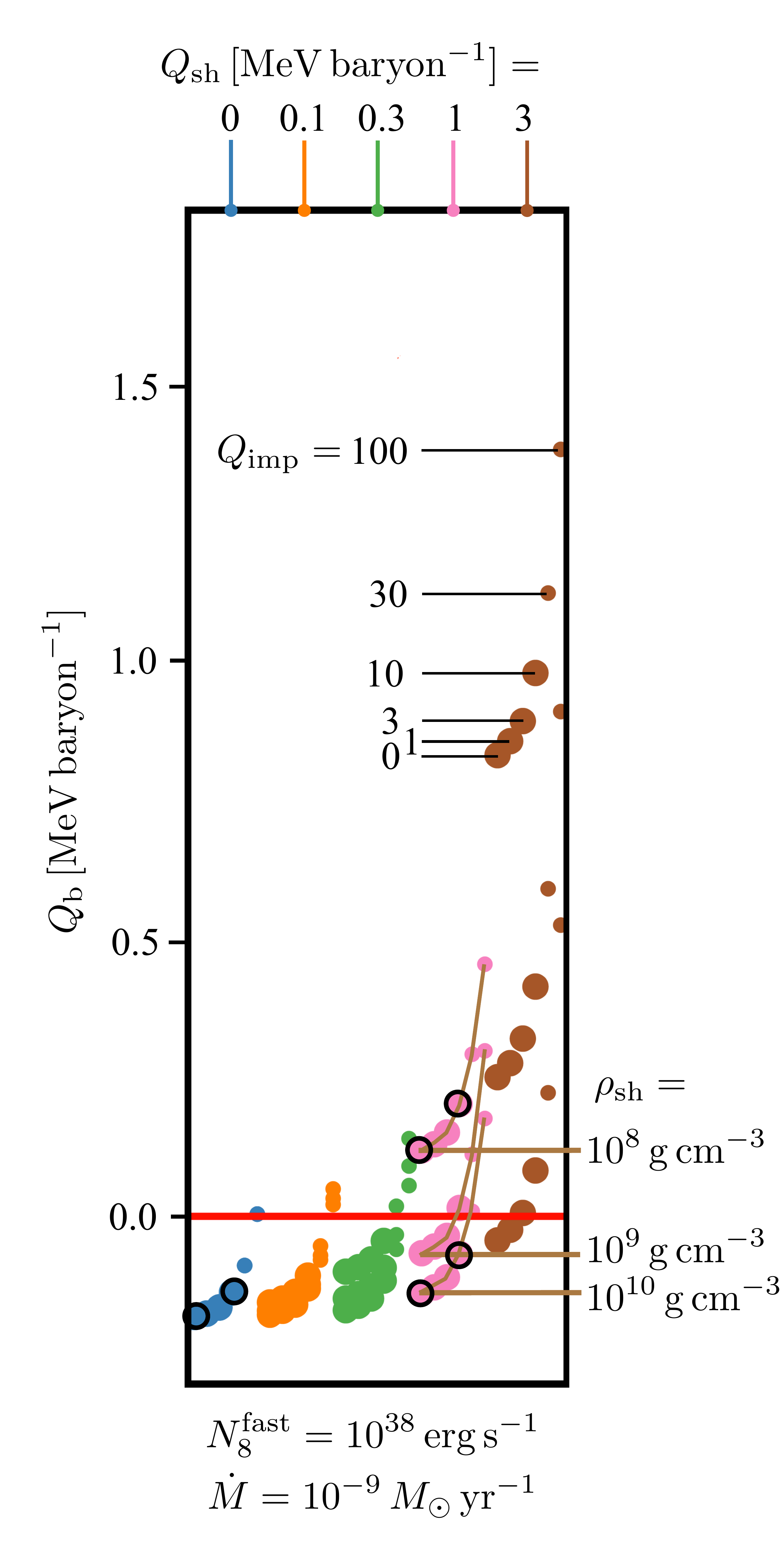}
\end{center}
 \caption{Left panel: temperature profiles inside neutron star models persistently accreting at a rate of $10^{-9} \, M_\odot$ yr$^{-1}$ and undergoing fast core neutrino cooling with $\log_{10} N_8^\mathrm{fast} \, [\mathrm{erg \, s}^{-1}] = 38$.
 Dots at $\rho = 10^7$ g cm$^{-3}$ mark the interface between the stationary envelope and the interior: the plot shows interior profiles from \texttt{NSCool} connected to an envelope profile similar to the ones of Figure \ref{fig:T-prof}.
Models A, B, and C, correspond to different amounts of shallow heating: 
 1 MeV baryon$^{-1}$ deposited in the density range $[10^8,10^9]$ g cm$^{-3}$ 
 (marked by a pastel red background) in A,
 and in the range $[10^{10},10^{11}]$ g cm$^{-3}$  (marked by a pastel blue background) in B, while C has no shallow heating.
 Models plotted as continuous lines have a uniform $Q_\mathrm{imp} = 10$ applied in the whole crust, while models plotted as dashed lines have no impurity added.
 The curves of the melting temperature, $T_\mathrm{m}$, corresponding to a Coulomb coupling parameter $\Gamma = 180$,
 the neutron drip line, and the crust-core transition are also marked. Right panel: zoom-in of the frame of Figure \ref{fig:Nu-Mdot} with the same values of $N^\mathrm{fast} = 10^{38}$ erg s$^{-1}$ and $Q_\mathrm{sh}= 1$ MeV/baryon as in the left panel. The cyan color circles single out the six models depicted in the left panel.
 }
\label{fig:analysis}
\end{figure*}

The result of major interest from this modeling is the luminosity at the crust-envelope interface, converted into heat per accreted baryon, $Q_\mathrm{b}$, see Equation (\ref{Eq:Lb-Tb}), which we present in Figure \ref{fig:Nu-Mdot} where one can see how it depends on the various parameters.
We are considering five values of $\log_{10} N_8^\mathrm{slow}$, nine values of
$\log_{10} N_8^\mathrm{fast}$, and six values of $\dot{M}$ giving us 84 combinations that correspond to the 84 frames of the figure.
Moreover, we explore five values of $Q_\mathrm{sh}$ resulting in five vertical series of points in each frame: each series contains $3$ strings of 6 points corresponding to three different densities $\rho_\mathrm{sh}$ and six values of $Q_\mathrm{imp}$.
Notice that we plot as large dots models with $Q_\mathrm{imp} \leq 10$, while models with very large $Q_\mathrm{imp}$, 30 or 100 that may be too large, are plotted as small dots.
In total we have 7560 models.

A cursory look at Figure \ref{fig:Nu-Mdot} shows that under slow neutrino cooling, left panel, in most cases heat flows from the crust into the envelope, i.e., $Q_\mathrm{b}>0$, with only a few exceptions: in all these cases the crust turns out to be hotter than the nuclear burning region of the envelope because the core itself is hot, having a low neutrino luminosity.
The few exceptions are for the highest values of $N^\mathrm{slow}_8 \geq 10^{34}$ erg s$^{-1}$ and not too high mass accretion rates
that result in a small inward heat flow $Q_\mathrm{b}$ of the order of 100 keV baryon$^{-1}$.

This situation changes significantly when considering the models with fast neutrino cooling (right panel): increasingly stronger neutrino cooling results in colder cores and crusts into which heat can flow from the envelope, i.e., $Q_\mathrm{b}<0$, in a large number of cases.
The number of models in which $Q_\mathrm{b}<0$ also has a strong dependence on the mass accretion rate and on the shallow heating:
larger values of either $Q_\mathrm{sh}$ or $\dot{M}$ result in hotter crusts that are more likely to be hotter than the nuclear burning envelope.

Notice that almost no models with a high accretion rate $\dot{M} > 3\times 10^{-9} \, M_\odot$ yr$^{-1}$, and only a small fraction of the ones with 
$\dot{M} = 3\times 10^{-9} \, M_\odot$ yr$^{-1}$, have a negative $Q_\mathrm{b}$, independently of the implemented $Q_\mathrm{sh}$: in all these situations we almost always have a strong heat flow from the crust into the envelope.
It is interesting that these mass accretion rates are precisely the ones where observationally X-ray bursts are seen to be quenched (e.g. \citet{Cornelisse:2003aa, Galloway:2008aa, Galloway:2021aa}).

It is worth noticing that the cases of slow and fast neutrino cooling with $N_8^* = 10^{35}$ erg s$^{-1}$ present almost identical results: this is simply due to the fact that stars with such neutrino luminosities all have core temperatures $\sim 10^8$ K so that the difference between the $T_8^6$ and $T_8^8$ dependence is negligible.

To better understand these results we show in Figure \ref{fig:analysis} several temperature profiles and a detailed view of a single frame from Figure \ref{fig:Nu-Mdot}.
In the left panel we display six temperature profiles with different values of the shallow heating, $Q_\mathrm{sh}$, and its depth of deposition, $\rho_\mathrm{sh}$, as well as different impurity contents, $Q_\mathrm{imp}$.
We single out the case of 
$N^\mathrm{fast}_8 = 10^{38}$ erg s$^{-1}$ and $\dot{M} = 10^{-9} M_\odot$ yr$^{-1}$ as it is very illustrative, but the other cases show exactly the same pattern.
The impact of reducing the thermal conductivity by increasing electron scattering with impurities is clear: the models with $Q_\mathrm{imp} = 10$ have much hotter crusts than models with $Q_\mathrm{imp} = 0$ simply because the heat generated in the crust, by both the deep crustal and the shallow heatings, is flowing more slowly into the core (or sometimes outward into the envelope, as in the models labeled ``A'').
With regards to the shallow heating, an obvious effect is that models with this heat source acting, as models A and B, have hotter crusts  than models where this heat source is not acting, as models C. 
More interesting, we see that when this source is located at lower densities, as in models A compared to models B, its impact on the temperature is much stronger: this is simply due to the fact that lower density matter has a lower specific heat and therefore becomes hotter for the same amount of injected heat.

The whole set of results for these selected values of $N^\mathrm{fast}_8$ and $\dot{M}$ are displayed in the right panel of
 Figure \ref{fig:analysis} that just reproduce the same frame from Figure \ref{fig:Nu-Mdot} and the six models shown in the left panel as singled out by cyan circles.
In the upper part of this panel values of $Q_\mathrm{imp}$ are indicated: this clearly shows that, all other parameters being equal, a lower conductivity from larger $Q_\mathrm{imp}$ results in a hotter crust that allows more heat to flow outward when 
$Q_\mathrm{b}>0$ or less heat to flow inward in the cases when $Q_\mathrm{b}<0$.
For each chosen value of $Q_\mathrm{sh}$ we can distinguish the three sub-series of points corresponding to the three chosen values of $\rho_\mathrm{sh}$: 
these three sub-strings are explicitly marked in the case $Q_\mathrm{sh}=1$ MeV baryon$^{-1}$ and all three exhibit the same clear pattern of  $Q_\mathrm{imp}$-dependence.
Moreover, a comparison of these three sub-strings clearly shows that a lower density $\rho_\mathrm{sh}$ results 
more heat to flow outward when  $Q_\mathrm{b}>0$ or less heat to flow inward in the cases when $Q_\mathrm{b}<0$:
this is exactly the same specific heat effect as noted in the difference between models A  and B in the left panel.
 
In view of these interpretations, returning to Figure \ref{fig:Nu-Mdot} we can still notice, as a curiosity, an interesting physical effect. When considering increasing values of $\dot{M}$, or increasing values of $Q_\mathrm{sh}$, i.e., increasing amounts of heating and thus increasingly hotter crusts, the sub-strings (corresponding to fixed values of $Q_\mathrm{sh}$ and $\rho_\mathrm{sh}$) that present the effect of $Q_\mathrm{imp}$ become less and less inclined: 
this reflects the fact that impurities have no effect in the liquid phase and that increasingly hotter crusts have a liquid outer region that reaches deeper into the crust, thus reducing the effect of impurities.
In the extreme case of the highest values of both $\dot{M}$ and $Q_\mathrm{sh}$ we see that the crust is so hot, for all possible neutrino luminosities, that the resulting $Q_\mathrm{b}$ have no dependence on $Q_\mathrm{imp}$ at all.

Finally, we note that under fast neutrino cooling and at mass accretion rates $\dot{M} < 3\times 10^{-9} \, M_\odot$ yr$^{-1}$ almost all models with shallow heating $Q_\mathrm{sh}$ smaller that 300 keV, and a large fraction of models with $Q_\mathrm{sh} = 1$ MeV, have heat inflow from the envelope into the crust unless $Q_\mathrm{imp}$ is extremely large, 30 or 100.

\section{Modeling Transiently Accreting Sources with their Thermonuclear Burning.}
\label{sec:Transient_Accretion}

\begin{figure*}
\begin{center}
\includegraphics[trim={0 0 0 1.0cm},width=0.90\textwidth, angle=0]{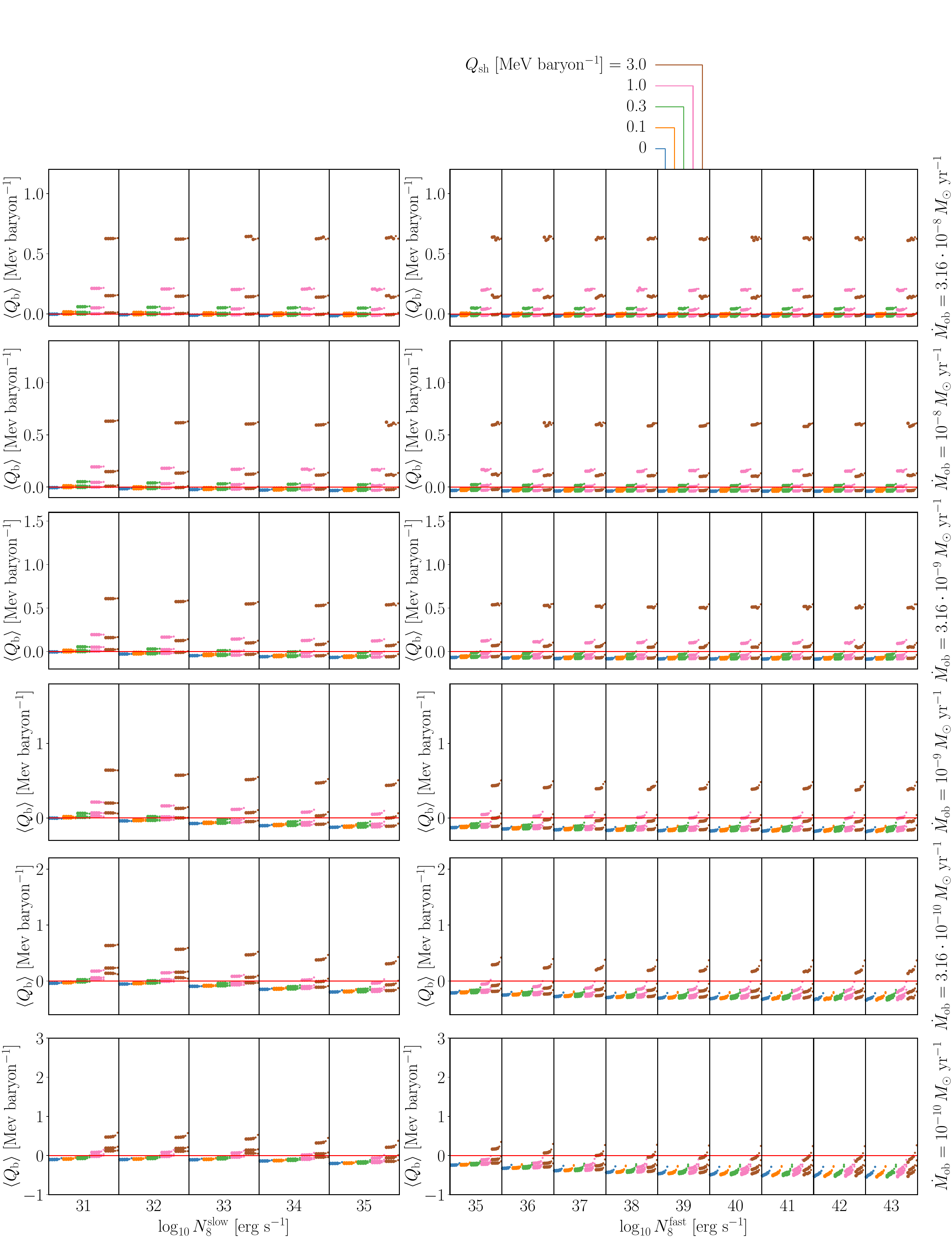}
\end{center}
 \caption{Average energy per accreted baryon, $\langle Q_\mathrm{b} \rangle$, transferred between the crust and the envelope as a function of the neutrino luminosity, $N^\mathrm{slow}_8$ for slow processes in the left panel and $N^\mathrm{fast}_8$ for fast processes in the right panel, which are, moreover, separated in six rows of frames according to the assumed mass accretion rate $\dot{M}_\mathrm{ob}$ during an outburst of one month duration for systems with a recurrence time of 10 months.
In each frame with given values of $N^*_8$ and $\dot{M}$, the five vertical series of points correspond to five different amounts of shallow heating, as indicated at the top of the figure, and each series consists of 3 strings of 6 points according to three different values of $\rho_\mathrm{sh} = 10^8$, $10^9$, and $10^{10}$ g cm$^{-3}$ and six values of $Q_\mathrm{imp} =$ 0, 1, 3, 10, 30, and 100. 
Details of a sample frame are described in Figure \ref{fig:analysis}.
Heat flows from the envelope into the crust when $Q_\mathrm{b}<0$.
}
\label{fig:Nu-Mdot-Transient}
\end{figure*}

We here present a first step in applying our accreting envelopes to transiently accreting neutron stars.
This is a direct extension of the method of the previous section with the intent to mimic the effect of recurrent outbursts of accretion separated by longer periods of quiescence.
Considering the idealized case of a perfectly periodic system undergoing outbursts with constant mass accretion rate $\dot{M}_\mathrm{ob}$, this introduces at least two more parameters, which are the length $t_\mathrm{ob}$ of these outbursts and the recurrence time between one outburst and the next, $t_\mathrm{rec}$.
For this first attempt, we will choose $t_\mathrm{ob} =$ 1 month and
$t_\mathrm{rec} = 10 \, t_\mathrm{ob}$, i.e., a 10\% duty cycle, which is roughly representative of several systems \citep{2024arXiv240718867H}.
Due to the very high heat capacity of the neutron star core its temperature does not change appreciably during such short outbursts \citep{PhysRevC.95.025806} and the effect of many recurrent outbursts is to bring this temperature to a value such that, over a long time, the average energy losses by neutrinos and photons balance the average heating from the accretion \citep{2001ApJ...548L.175C}.

In view of this we proceed as follow:
we apply a constant average mass accretion rate $\langle \dot{M} \rangle$ that is 10\% of $\dot{M}_\mathrm{ob}$, running the simulation for a million years for the core to reach this long term equilibrium, then we stop accretion for five years to let the crust cool and reach thermal equilibrium with the core, resulting in a star with an isothermal interior, and then turn on accretion at $\dot{M}_\mathrm{ob}$ to simulate an outburst.
During this outburst we register the amount of heat transferred at $\rho_\mathrm{b}$ and deduce an average $\langle Q_\mathrm{b} \rangle$.
At the start of accretion the burning envelope rapidly heats up and $Q_\mathrm{b}$
is negative in most cases, but, as time passes, the crust slowly heats up and $Q_\mathrm{b}$ may turn positive or remain negative depending on the details of the model.
We consider the same model parameters as in the case of persistent accretion: $\dot{M}$, now considered as being $\dot{M}_\mathrm{ob}$, the core neutrino luminosities, $N_8^*$, the shallow heating strength, $Q_\mathrm{sh}$, and its depth, $\rho_\mathrm{sh}$, and the amount of impurity in the crust, $Q_\mathrm{imp}$.

\subsection{Results.}
\label{sec:Transient_Accretion_Results}

We summarize the results of these simulations in Figure \ref{fig:Nu-Mdot-Transient} using the same presentation as in Figure \ref{fig:Nu-Mdot}, but emphasize that we now present the values of the average $\langle Q_\mathrm{b} \rangle$ for a one month-long accretion outburst.
Overall, this figure has a similar appearance to Figure \ref{fig:Nu-Mdot}, but one clearly sees that there are many more cases with heat flowing from the envelope into the crust, i.e., with $\langle Q_\mathrm{b} \rangle < 0$.
When considering longer outbursts, but with the same duty cycle of 10\%, very similar outcomes are obtained.
Longer outburst with larger duty cycles will become increasingly closer to the persistent cases.

One interesting feature is worth noting: one sees that, in most cases, when $Q_\mathrm{sh}$ is deposited at the highest density, the values of $\langle Q_\mathrm{b} \rangle$ are independent of $Q_\mathrm{imp}$, $N_8$, and $Q_\mathrm{sh}$. 
The reason for this is simply that the duration of the accretion outburst is shorter than the thermal time scale at the deposition density of $10^{10}$ \gcc, while the effect of nuclear burning at low densities is almost immediate.
In other words, the short time scale evolution of the low density regions is naturally decoupled from the slowly evolving denser regions.
On the other hand, one sees that the effect of impurities, that is almost nil at high $\dot{M}$, increases at lower $\dot{M}$, again due to the fact that hotter crusts have a deeper liquid ocean where impurities play no role.

\begin{figure*}
\begin{interactive}{animation}{Movie_AB.mp4}
\centerline{\includegraphics[width=0.75\textwidth]{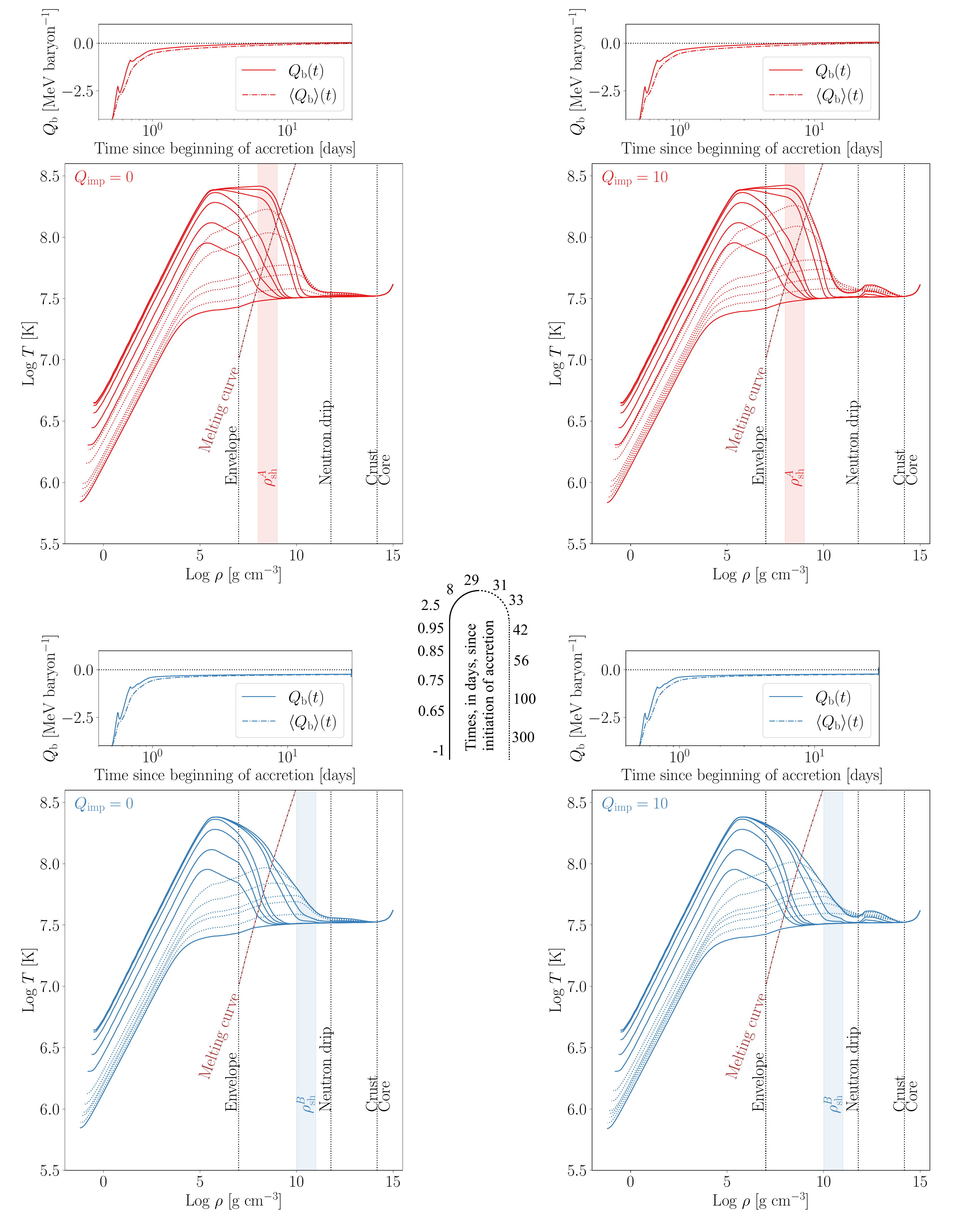}}
\end{interactive}
\caption{Time evolution of the temperature profile for four models across a 30 day accretion outburst at $\dot{M}_\mathrm{ob} = 10^{-9} \, M_\odot \, \mathrm{yr}^{-1}$ in systems with a recurrence time of 300 days.
Profiles before, at day -1, and during the accretion outburst up to day 29 are plotted as continuous lines, while profiles during the following quiescent phase, from day 31, are plotted as dotted lines.
In each case, the time evolution during the accretion phase of the instantaneous $Q_\mathrm{b}(t)$, as well as the building of the average $\langle Q_\mathrm{b}\rangle (t)$ taken from $t'=0$ to $t$, are shown in the upper inset.
Models of the two upper (lower) panels have the same physics as the models A (B) of Figure \ref{fig:analysis}, with $Q_\mathrm{imp} = 0$ in the left panels and $Q_\mathrm{imp} = 10$ in the right ones.
In the left panels, the last profile, at day 300, is not discernible as it coincides with the initial profile at day -1.
Numerically, $\dot{M}$ is increased exponentially during the first day of the outburst till it reaches $\dot{M}_\mathrm{ob}$ and decreased exponentially during the last day till it reaches 0 at day 30.
The two peaks in $Q_\mathrm{b}(t)$ at early times are numerical artifacts as the early envelope models at very low $\dot{M}$ are extrapolations of the ones calculated in Section \ref{sec:stat_td_envelopes}.
\\
An animated version of this figure is available in the on-line version of this paper: it presents the time evolution, from day -1 to day +300, of the temperature profiles displayed in the four panels of the figure.}
\label{fig:analysis2}
\end{figure*}

To further illustrate these results we show in Figure \ref{fig:analysis2} the time evolution of the temperature profile during the accretion heating phase and the subsequent cooling during quiescence, as well as the time evolution of $Q_\mathrm{b}(t)$, in four cases similar to the ones of Figure \ref{fig:analysis}.
In all four cases one first notices the rapid heating, by nuclear energy, of the envelope at the beginning of the accretion phase and the later diffusion of this heat inward as the envelope is much hotter that its underlying layers.
We emphasize that the first few hours of evolution at the start of the outburst cannot be followed accurately with our models of stationary envelopes and such a study should be performed with time-dependent models as presented in Figure \ref{fig:mesa_different_rates}.
The dramatic effect of the depth of deposition of the shallow heating is clearly seen when comparing the two upper panel models with the ones in the lower panels.
In the upper panels the shallow heating  is acting at low densities, $\rho_\mathrm{sh} = 10^8$ \gcc, inducing a strong increase in temperature of these layers, that reduces the temperature gradient with the envelope and eventually results in an inversion of the heat flow, now from the crust into the envelope.
In contrast, in the lower panel models with shallow heating located deeper, $\rho_\mathrm{sh} = 10^{10}$ \gcc, the effect is seen to be almost negligible: matter at these densities has a large enough specific heat that it requires a much longer accretion phase to see its temperature rise significantly.
In these latter cases there will always be a inward temperature gradient during the accretion phase and flow of thermonuclear energy into the crust.
Complementarily, the effect of impurities in reducing the thermal conductivity in the solid phase when the temperature $T$ is much below the melting temperature $T_\mathrm{m}$ is illustrated by comparing the two left panel models, $Q_\mathrm{imp}=0$, with the two right panel ones, $Q_\mathrm{imp}=10$.
During the heating phase and in the low density regions, e.g. below $10^{10}$ \gcc, there is no noticeable difference between the models of the left and right panels: impurities play no role as matter is either in a liquid state or a solid with $T$ not too much below $T_\mathrm{m}$.
In contrast, in the deeper layers of the inner crust, where the pycnonuclear reactions are occurring, in models with high impurity the heat released by these reactions is stored for a longer time resulting in a temperature bump, while in the low impurity models it can efficiently diffuse inward into the core resulting in a much smoother temperature profile in the inner crust.
At later times, during the cooling phase, the effect of impurities is felt within the whole crust as the evolution of the outer crust is now strongly coupled to the evolution of the inner crust: after 300 days of evolution the temperature profiles of the low impurity models have returned to their pre-outburst values, whereas in the cases of the high impurity models the whole crust is now warmer than it was at the beginning of the outburst. 
This effect is most clearly seen in the animated version of the figure and was explicitly described based on observations of the Aql X-1 system by \citealt{Ootes:2018aa}.

\section{Discussion and Conclusions}
\label{sec:Conclusions}

Our interest here has been to improve our understanding of the heating of a neutron star during outbursts of accretion.
We focused on neutron stars in low-mass X-ray binaries as they present an extremely rich phenomenology, can undergo persistent or transient accretion, the nuclear burning of the accreted matter on their surface can proceed stably at high enough mass accretion rates or become unstable and produce X-ray bursts and superbursts, and, moreover, they have very small magnetic fields that have little or no effect on most of their observed characteristics.
We have focused on one source of energy that has often been considered as having no effect on the interior evolution of the star:
the thermonuclear reactions occurring in the surface ocean.
Most authors assume that the neutron star crust is hotter than the burning envelope and that there is always heat flowing into the envelope from the crust.
We have presented models of accreting neutron star envelopes, either in a stationary state using our own envelope code \citep{2024arXiv240313994N}, or in time dependent states, using the public code \texttt{MESA}  \citep{Paxton:2011aa,Paxton:2013aa,Paxton:2015aa,Paxton:2018aa,Paxton:2019aa,Jermyn:2023aa},
in which the nuclear burning layer is hotter than the bottom of the envelope and in which heat is thus flowing from the envelope into the neutron star crust.
We here notice that \citet{2009A&A...497..469I} found some observational evidence of inward heat flux during X-ray bursts in the long burst tails of GS 1826--24.

A self-consistent evaluation of the possibility of having thermonuclear energy flowing into the stellar interior from the burning envelope requires modeling the whole star temperature evolution, i.e., the coupling between all possible energy sources and energy sinks (mostly neutrinos), over possibly long time scales, to determine in which cases the burning envelope is hotter or colder that the underling crust.
Such models have already been described 
\citep{1984PASJ...36..199H,1984ApJ...278..813F,Fujimoto1987op,Miralda-Escude1990wd,Zdunik1992bn,Dohi2021, Dohi2022yu} under different conditions and assumptions. They clearly showed the possibility of an inward energy flux in the presence of very strong neutrino emission in the stellar core.
\citet{1984PASJ...36..199H} and \citet{Dohi2021, Dohi2022yu} followed in details the unstable burning, resulting in X-ray bursts, but could only study the evolution of the star for a few hours/days, as such calculations are extremely time consuming, while the other authors implemented stationary envelope models and could thus study the long term evolution of the star.
To provide a stronger theoretical basis for the long term study of accreting neutron stars we showed, in Section \ref{sec:stat_td_envelopes}, that stationary envelope models provide a reasonable first approximation to time-dependent envelope models when focusing on the amount of heat $Q_\mathrm{b}$ transferred between the crust and the envelope.
When performing such evolutionary calculations the envelope simply provides a relationship between the temperature at the bottom of the envelope, $T_\mathrm{b}$, and the luminosity at that point, $L_\mathrm{b}$, or equivalently $Q_\mathrm{b}$ from Equation (\ref{Eq:Lb-Tb}), that is used as an outer boundary condition for the modeling of the interior.
We could provide an extensive set of such models, whose results are displayed in Figure \ref{fig:Lb-Tb}, covering a wide range of mass accretion rates and boundary temperatures.
The right panel of this figure shows that values of $Q_\mathrm{b}$ of more than one MeV baryon$^{-1}$ are possible in the case of a cold crust.

Armed with this extensive set of boundary conditions, we modeled the whole thermal evolution of neutron stars undergoing either persistent or transient accretion using an updated version our public neutron star cooling code \texttt{NSCool} \citep{NSCool}.
The results, presented in Figures \ref{fig:Nu-Mdot} and \ref{fig:Nu-Mdot-Transient}, show which combinations of neutrino luminosity and mass accretion rate lead to a heat flow from the crust into the envelope, $Q_\mathrm{b} >0$, or from the envelope into the crust,  $Q_\mathrm{b} <0$.
High mass accretion rates $\dot{M}$ and/or low neutrino luminosities $L_\nu$ result in hot interiors and positive $Q_\mathrm{b}$, while lower $\dot{M}$ and/or higher $L_\nu$ result in colder interiors and negative $Q_\mathrm{b}$.
The direction of the heat flow at the envelope-crust interface is moreover very sensitive to the thermal conductivity of the crust and the presence, as well as location, of shallow heating. 
A low crust thermal conductivity, due to a large impurity content $Q_\mathrm{imp}$, keeps the crust warmer and makes it more unlikely to have an inward energy flow from the envelope into the crust.
Similarly, a strong shallow heating $Q_\mathrm{sh}$ also keeps the crust warmer and has a similar effect, particularly when the energy deposition density 
$\rho_\mathrm{sh}$ is low, close to the envelope-crust interface.
This interplay between conductivity and shallow heating is illustrated  in detail in Figure \ref{fig:analysis}.
When describing systems with transient accretion, that have colder cores than persistent sources due to the periods of quiescence, we found that they are more likely to have inward fluxes.

We find that, when there is heating of the crust by leakage of thermonuclear energy from the burning envelope, negative values of  $Q_\mathrm{b}$ of at most a few hundreds keVs are obtained in the most extreme cases, while the results of Figure \ref{fig:Lb-Tb} indicated that much higher values could be expected.
The reason for this reduction is clearly illustrated by the evolution of the temperature profiles in the case of transient accretion that are displayed in Figure  \ref{fig:analysis2}.
This figure describes the beginning of an accretion outburst in the case of a cold star and shows that initially, during the first day, there is actually a very strong heat flow from the envelope into the crust.
However, the layers just below the envelope have a relatively small specific heat and, thus, see their temperature increase rapidly under the strong heating.
As a result, the inward temperature gradient is strongly reduced and the heat flow is quenched.
These results seem to strongly indicate that the shallow heating, which has been found to require energies of the order of a few MeV baryon$^{-1}$
\citep{Brown:2009aa,Page2013ty,Degenaar2014io,Turlione2015rt,Degenaar2015nm,Merritt2016,Parikh2018kl,Parikh2019,Ootes2019rt,Degenaar2021bn}, cannot be described by the leakage of thermonuclear energy from the envelope, but detailed studies of each system with all its specific characteristics are necessary to confirm this conclusion.

Finally, an important by-product of our extensive study is our description of the correlation between the amount of shallow heating, together with the depth of its deposition, and the amount of heat flowing into the burning envelope in the cases where $Q_\mathrm{b}$ is positive. This can be seen in Figures \ref{fig:Nu-Mdot} and \ref{fig:Nu-Mdot-Transient}, together with its dependence on the various factors $\dot{M}$, $L_\nu$ and $Q_\mathrm{imp}$.
A strong heat flow from the crust has been proposed as the mechanism for the quenching of X-ray bursts at high mass accretion rates 
\citep[e.g.][]{keek2009,Zamfir:2014aa,2016MNRAS.456L..11K} and our results can allow to infer the properties of the neutron star crust, e.g. $Q_\mathrm{imp}$, and core, e.g. $L_\nu$, required to provide the needed amount of heating. We note that alternative explanations for the quenching of the bursts exist. Some still include the effects of extra heating, even if not related to flux from the crust, for example from magnetohydrodynamical processes \citep{Inogamov1999,Inogamov2010,piro2007,keek2009}. Others propose that the quenching is related to different effective local accretion rates at different latitudes \citep{1998ASIC..515..419B,Cooper2007,Cavecchi2017,Cavecchi:2020aa}. These alternative explanations do not exclude a flux from the crust into envelope, although they may reduce the amount required, and can still be coupled with our results to put constraints on the conditions in the crust and core.

What is missing in our present study is an evaluation of the effect of the star mass and radius.
These impact on all radial length-scales and will alter the details of our results. 
For example, \citet{1982ApJ...259L..19G} showed that $L_\mathrm{b}$ is directly proportional to the gravitational acceleration in the envelope.
Also of interest will be the dependence on the fraction of H/He in the accreted matter that has a strong impact on the nuclear burning.
These issues will be addressed in a future paper.

\section*{Acknowledgments}
M.N.-C. acknowledge supports from a fellowship of Conahcyt. His work was also supported by the Fonds de la Recherche Scientifique-FNRS under Grant No IISN 4.4502.19. His and D.P.'s work is also supported by a UNAM-DGAPA grant PAPIIT-IN114424.
Y.C. acknowledges support from the grant RYC2021-032718-I, financed by MCIN/AEI/10.13039/501100011033 and the European Union NextGenerationEU/PRTR.
We thank the referee for useful suggestions which helped improving this manuscript.

The inlist for the \texttt{MESA} simulations in this work can be found at \cite{nava_callejas_2025_15828636}


\bibliographystyle{aasjournal}
\bibliography{Envelope_Negative_Lb}

\appendix

\section{On the relativistic corrections to \texttt{MESA}}\label{appendx:gr_corr_1}

\texttt{MESA} allows to handle some general relativistic scenarios, such as neutron star envelopes, by multiplying the universal constant of gravitation $G$ with a ``correction'' factor $c_{\text{grav}}$, i.e. $G \to G c_{\text{grav}}$ with $c_{\text{grav}} := e^{2\Lambda}\frac{\mathcal{H}}{\rhobaryon}\mathcal{G}$. Here, $\rhobaryon$ is the rest-mass density, $e^{2\Lambda} = \left[1 - \frac{2Gm}{c^{2}r}\right]^{-1}$ is the metric function and $\mathcal{H} = \rhobaryon + P/c^{2}$, $\mathcal{G} = 1 + \frac{4\pi r^{3}P}{mc^{2}}$. While this treatment is appropriate for the structure equations, it yields an incomplete transformation for the energy transport:
\begin{itemize}
	\item \textbf{Temperature gradient - radiative case.} The non-relativistic equation reads
	\begin{equation}
	\nabla_{T} = \frac{3LP\kappa}{64\pi\sigma_{\text{SB}}T^{4}Gm}.
	\end{equation}
	Upon $G\to G c_{\text{grav}}$ replacement, 
	\begin{equation}
	\nabla_{T}[\texttt{MESA}, \text{cgrav}] = \frac{3LP\kappa\rhobaryon}{64\pi\sigma_{\text{SB}}T^{4}Gme^{2\Lambda}\mathcal{H}\mathcal{G}}.
	\end{equation}
	On the other hand, in a fully general-relativistic treatment we obtain \citep{Thorne_1977ApJ...212..825T}:
	\begin{align}
	\nabla_{T}[\text{GR}] & = \frac{3LP\kappa\rhobaryon}{64\pi\sigma_{\text{SB}}T^{4}Gme^{\Lambda}\mathcal{H}\mathcal{G}} + \left(1-\frac{\rho}{\mathcal{H}}\right) \approx \frac{3LP\kappa\rhobaryon}{64\pi\sigma_{\text{SB}}T^{4}Gme^{\Lambda}\mathcal{H}\mathcal{G}}
	\end{align}
	due to $\rho\approx\mathcal{H}$ in systems such as the envelope. Therefore
	\begin{equation}
	\nabla_{T}[\texttt{MESA}, \text{cgrav}] \approx e^{-\Lambda}\nabla_{T}[\text{GR}].
	\end{equation}
	%
%
	\item \textbf{Luminosity equation.}\footnote{Recall \texttt{MESA} solves $dL/dm$. However, this discussion illustrates how the $dL/dP$ version in \texttt{MESA} looks like in order to perform a fair comparison.} In standard stellar structure, starting from $dL/dm =$ $\sum^{3}_{j=1}\dot{\breve{\varepsilon}}_{j}$ and upon repeated applications of the chain rule, we arrive to
	\begin{equation}
	\frac{dL}{dP} = -\frac{4\pi r^{4}}{Gm}\sum^{3}_{j=1}\dot{\breve{\varepsilon}}_{j}.
	\end{equation}
	Thus, upon $G\to Gc_{\text{grav}}$ replacement we arrive to
	\begin{equation}
	\frac{dL}{dP}[\texttt{MESA}, \text{cgrav}] = -\frac{4\pi r^{4}\rhobaryon}{Gme^{2\Lambda}\mathcal{H}\mathcal{G}}\sum^{3}_{j=1}\dot{\breve{\varepsilon}}_{j}.
	\end{equation}
	On the other hand, in a fully relativistic discussion we have 
	\begin{align}
	e^{-2\Phi}\frac{d(e^{2\Phi}L)}{dP}[\text{GR}] & = -\frac{4\pi r^{4}\rhobaryon}{Gme^{\Lambda}\mathcal{H}\mathcal{G}}\sum^{3}_{j=1}\dot{\breve{\varepsilon}}_{j}\\
	\frac{dL}{dP}[\text{GR}] & \approx -\frac{4\pi r^{4}\rhobaryon}{Gme^{\Lambda}\mathcal{H}\mathcal{G}}\sum^{3}_{j=1}\dot{\breve{\varepsilon}}_{j}.
	\end{align}
    given that variations of $e^{-\Phi}$ are small in the envelope. Thus
	\begin{equation}
	\frac{dL}{dP}[\texttt{MESA}, \text{cgrav}] \approx e^{-\Lambda}\frac{dL}{dP}[\text{GR}].
	\end{equation}
	%
%
\end{itemize}
From these arguments, we observe that proper relativistic corrections require further modifications of the underlying \texttt{MESA} equations. To simplify the process of comparison, we thus opted for neglecting the general-relativistic corrections in \texttt{MESA} and adapt our numerical code to produce the corresponding non-relativistic envelope models.

\clearpage
\section{Supplementary Material}\label{Suppl.}

\noindent
{\bf Variants of the persistently accreting cases}

\begin{figure*}
\begin{center}
\includegraphics[trim={0 0 0 1.0cm},width=0.90\textwidth, angle=0]{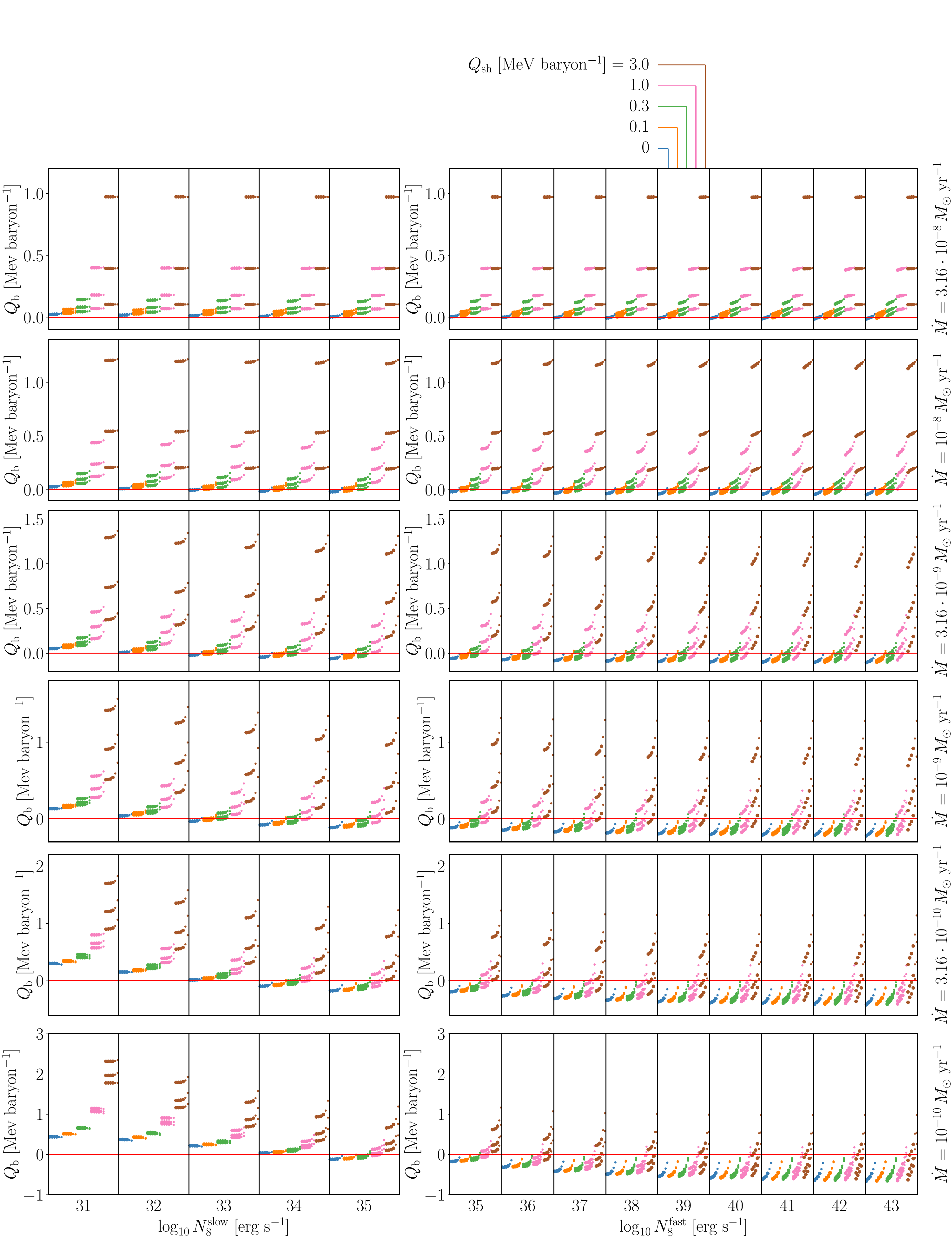}
\end{center}
 \caption{Same as Figure \ref{fig:Nu-Mdot} but suppressed deep crustal heating according to the model of \cite{2008A&A...480..459H}.
   }\label{fig:Nu-Mdot_2d}
\end{figure*}

In our study of persistently accreting systems in Section \ref{sec:Persistent_Accretion} we employed the model of \cite{2008A&A...480..459H} for the deep crustal heating that injects about 1.9 MeV baryon$^{-1}$, distributed in the crust with most of it released in the inner crust by the pycnonuclear fusions.
In the alternate model of \citet{Gusakov2020av} the energy released by the deep crustal heating is much smaller, of the order of 0.5 MeV and most of it is released in the outer crust with almost no heating occurring in the inner crust \citep{Potekhin2023fg}.
To evaluate the impact of this dramatic change in the inner crust physics on the heating of the outer crust by the envelope thermonuclear reactions we repeated the calculation leading to 
Figure \ref{fig:Nu-Mdot} which we present in Figure \ref{fig:Nu-Mdot_2d}.
Comparing these two figures one find that the effect is small, particularly in the cases where heat is flowing from the envelope into the crust.
In the opposite case, where heat flows from the crust into the envelope, one finds it can be less intense when the deep crustal heating is strongly suppressed in the inner crust.

\begin{figure*}
\begin{center}
\includegraphics[trim={0 0 0 1.0cm},width=0.90\textwidth, angle=0]{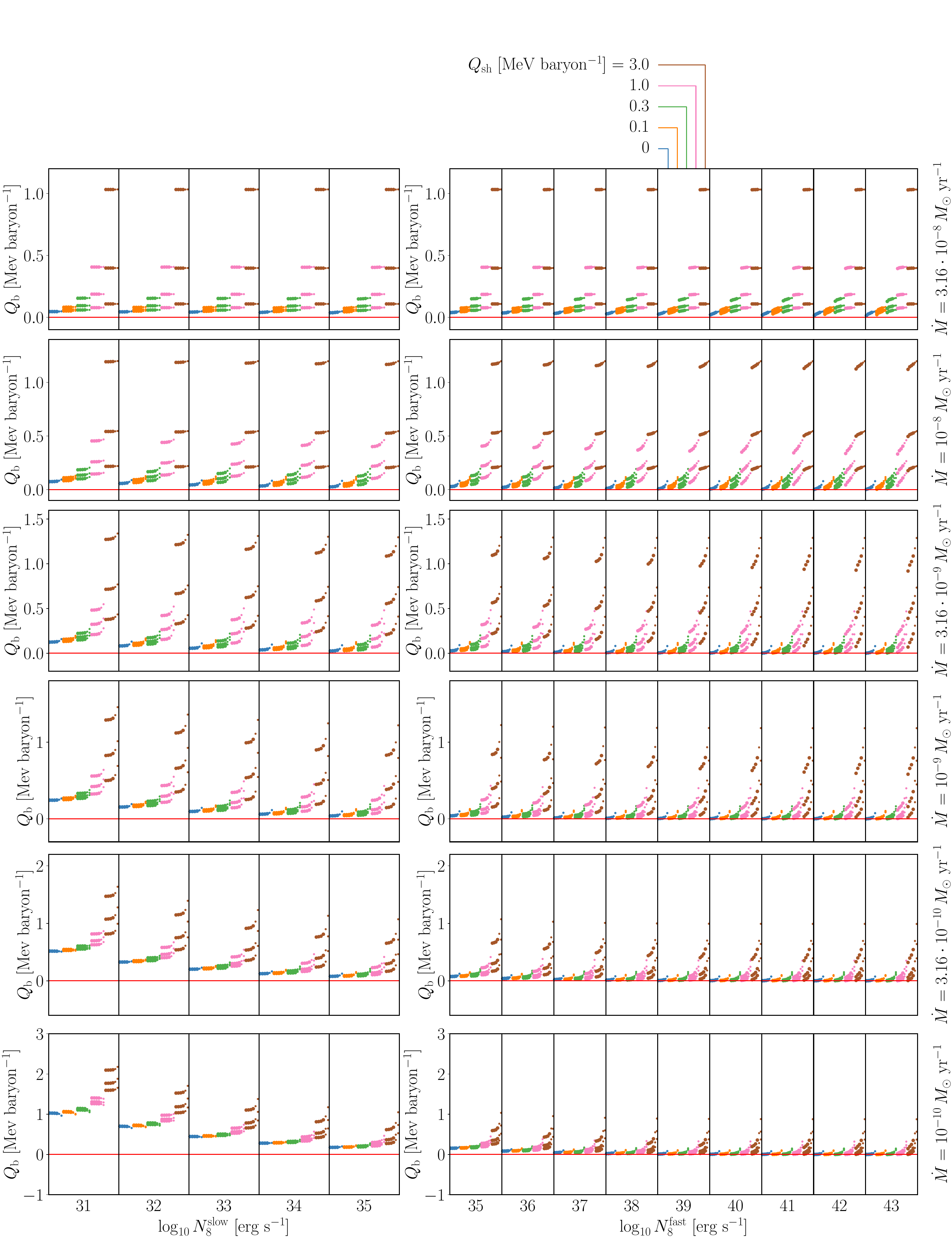}
\end{center}
 \caption{Same as Figure \ref{fig:Nu-Mdot} but with passively cooling envelopes that have no nuclear reactions acting.
   }
\label{fig:Nu-Mdot_2c}
\end{figure*}

Since our results are also of interest in quantifying the amount of heat injected into the envelope in the presence of shallow heating we also, for curiosity, repeated the same calculation but using envelope models in which no nuclear reaction occurs.
The results are presented in Figure \ref{fig:Nu-Mdot_2c}.
Naturally, in this case heat only flows from the crust into the envelope, but not having nuclear fusions imply lower temperature in the envelope and, thus, a strong temperature gradient in it which may drive a stronger heat flux, increasing $Q_\mathrm{b}$. 
However, comparison of Figure \ref{fig:Nu-Mdot} and Figure \ref{fig:Nu-Mdot_2c} shows that this possible effect is very small.

\noindent
{\bf Variants of the transiently accreting cases}

\begin{figure*}
\begin{center}
\includegraphics[trim={0 0 0 1.0cm},width=0.90\textwidth, angle=0]{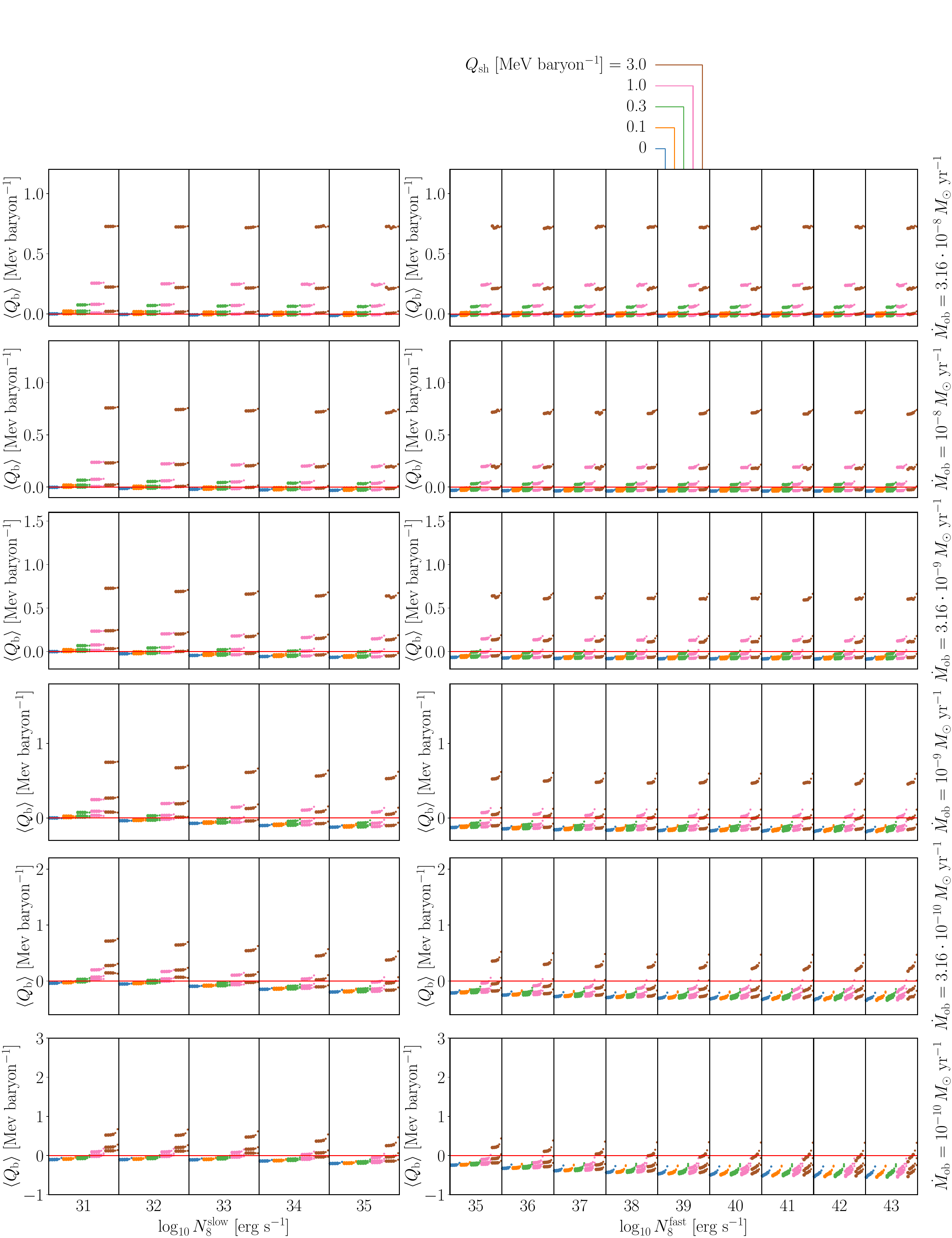}
\end{center}
 \caption{Same as Figure \ref{fig:Nu-Mdot-Transient} but with 2 months long accretion outburst.
   }
\label{fig:Nu-Mdot-Transient_2M}
\end{figure*}

\begin{figure*}
\begin{center}
\includegraphics[trim={0 0 0 1.0cm},width=0.90\textwidth, angle=0]{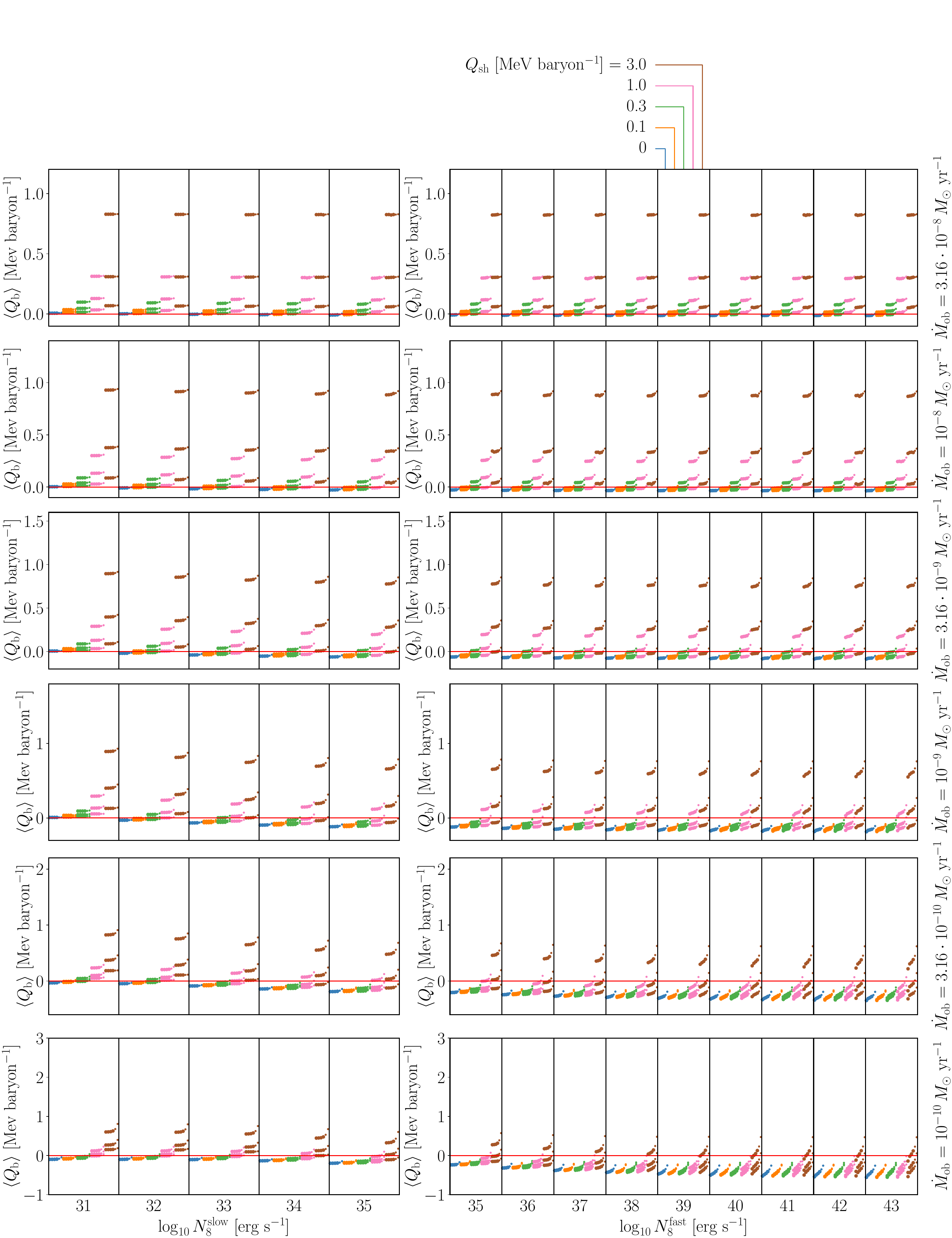}
\end{center}
 \caption{Same as Figure \ref{fig:Nu-Mdot-Transient} but with 1 year long accretion outburst.
   }
\label{fig:Nu-Mdot-Transient_1Y}
\end{figure*}

\begin{figure*}
\begin{center}
\includegraphics[trim={0 0 0 1.0cm},width=0.90\textwidth, angle=0]{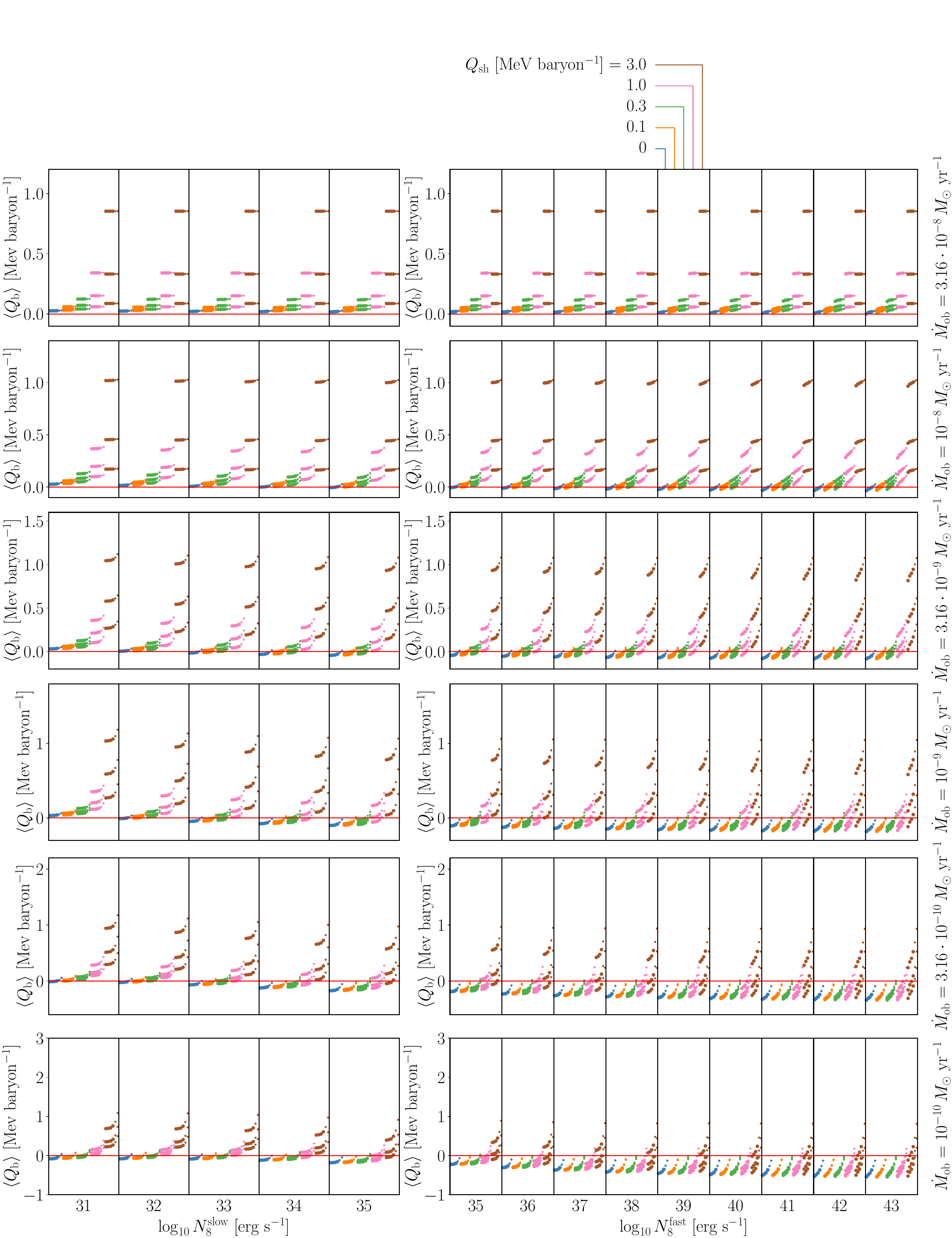}
\end{center}
 \caption{Same as Figure \ref{fig:Nu-Mdot-Transient} but with 1 century long accretion outburst.
   }
\label{fig:Nu-Mdot-Transient_1C}
\end{figure*}

We present here alternate figures to the results presented in Figure \ref {fig:Nu-Mdot-Transient} of Section \ref{sec:Transient_Accretion} for transiently accreting systems with longer outbursts, either 2 months in Figure \ref{fig:Nu-Mdot-Transient_2M}, one year in Figure \ref{fig:Nu-Mdot-Transient_1Y}, and one century in Figure \ref{fig:Nu-Mdot-Transient_1C}, but we keep a duty cycle of 10\%.
The results are very similar to the cases of one month of accretion.
Notice, however, that even the century long outbursts are still different than the cases of persistent accretion: 
this is due to the duty cycle of 10\% that results in a colder core than in the case of persistent accretion.

\end{document}